\documentclass{article}

\usepackage{xcolor}
\usepackage{booktabs}
\usepackage{musicography}
\usepackage{makecell}
\usepackage{comment}
\usepackage{float}

\newcommand{\specialcell}[2][c]{%
  \begin{tabular}[#1]{@{}c@{}}#2\end{tabular}}

\usepackage[utf8]{inputenc}
\usepackage[]{tismir}
\usepackage{amsmath}
\usepackage{hyperref}
\usepackage{url}
\usepackage{graphicx}
\usepackage{booktabs}
\usepackage{lipsum}

\title{The MPB Corpus: A Dataset of Melody, Rhythm, Harmony, and Melody-Harmony Relationships in Brazilian Popular Music}
\author{%
Carlos de L. Almada\thanks{School of Music, Federal University of Rio de Janeiro, Rio de Janeiro, Brazil},%
~Hugo T. de Carvalho\thanks{Department of Statistical Methods, Institute of Mathematics, Federal University of Rio de Janeiro, Rio de Janeiro, Brazil},%
~and Felipe D. Martins\footnotemark[1]~~\thanks{Institut für Theorie und Geschichte, Anton Bruckner Universität, Linz, Austria.}}

\date{}

\type{dataset}

\authorref{Almada,~C., Carvalho,~H., and Martins,~F.}
\authorshort{Almada, Carvalho and Martins} %or, e.g., \authorshort{Author1 et al}
\titleshort{The MPB Corpus}

\begin{document}

%%%%%%%%%%%%%%%%%%%%%%%%%%%%%%%%%%%%%%%%%%%%%%%%%%%%%%%%%%%%%%%%%%%%%%%%%%%%%%%%
% Abstract
%%%%%%%%%%%%%%%%%%%%%%%%%%%%%%%%%%%%%%%%%%%%%%%%%%%%%%%%%%%%%%%%%%%%%%%%%%%%%%%%

\twocolumn[{%
\maketitleblock
\begin{abstract}
\begin{center}
{\color{red} \center PLEASE CHECK AND REFER TO THE PUBLISHED VERSION OF THIS PAPER ON  \url{https://transactions.ismir.net/articles/10.5334/tismir.357}}
\end{center}
This paper presents the \textit{MPB Corpus}, a dataset of Brazilian Popular Music (\textit{Música Popular Brasileira}, MPB) comprising derived symbolic annotations for 500 pieces across four musical parameters -- melodic contour, melodic rhythm, harmony, and the melody–harmony relationship -- making it the most comprehensive dataset to date for computational musicology of Brazilian music. It is released under a CC-BY-4.0 license in an open, machine-readable tabular format, with documentation and reproducible analysis code. Because the annotations were derived from copyrighted musical scores, the corpus contains no audio recordings or score-level representations such as MIDI or MusicXML; it nonetheless provides a reusable resource for MIR research on a culturally significant repertoire that remains underrepresented in existing datasets. To support the encoding process, we introduce analytical models for capturing rhythmic and melodic information, along with tailored visualizations and summary metrics. Finally, we present a brief exploratory analysis illustrating the dataset's potential for formulating and addressing musicological questions about the genre.
\end{abstract}

}]
\saythanks{}

%%%%%%%%%%%%%%%%%%%%%%%%%%%%%%%%%%%%%%%%%%%%%%%%%%%%%%%%%%%%%%%%%%%%%%%%%%%%%%%%
% Main Content Start
%%%%%%%%%%%%%%%%%%%%%%%%%%%%%%%%%%%%%%%%%%%%%%%%%%%%%%%%%%%%%%%%%%%%%%%%%%%%%%%%

\section{Introduction}
\label{sec:introduction}
Corpus studies have emerged as a major trend in systematic musicology \citep{huron-big-data,huron-error-categories,white-music-data}, driven by advances in computational technology and by the expansion of large music databases of scores, audio recordings, and MIDI files. These studies span a broad range of repertoires and styles, each typically structured around working hypotheses, theoretical models, and methodologies tailored to its context and supported by statistical frameworks.

In recent years, corpus studies on popular music have become increasingly prevalent, covering a variety of genres and styles \citep{declercq-temperley-2011, mauch-etal-2015, serra-etal-2012, white-quinn-2018, hamilton-2024, hunke-2025}. Important as these references are in such a recent field, all but the last two are centered on harmony and correlated aspects, and the vast majority deal almost exclusively with anglophone popular music (mostly rock and pop, though some include jazz and blues). Notable exceptions focused on Latin American music are \cite{colombian-bambuco, candombe, salsa}, on Colombian \textit{bambuco}, Uruguayan \textit{candombe}, and \textit{salsa}. On Brazilian music, \cite{moss-etal-choro-2020, moss-choro} address \textit{choro}, but only harmonically; \cite{sambaset, brid, lucas-tismir} address \textit{samba} and \cite{maracatu-1, maracatu-2} \textit{maracatu de baque solto}, but only rhythmically. The present work aligns with this stylistic trend while considering not a single musical dimension but melodic and rhythmic structures, harmony, and the relation between the two. Moreover, our focus is one of the richest temporal-esthetic contexts of Brazilian music, better known by the acronym MPB. 

Although MPB literally means ``Brazilian Popular Music'' in Brazilian Portuguese (\textit{música popular brasileira}), the term refers neither to all popular music composed in the country nor exclusively to Brazilian popular \textit{urban} music (which emerged in Brazil around the 1850s), but to a specific subset of that vast production. The concept is in fact a cultural construct shared by Brazilians, and not easy to explain to non-natives.\endnote{For historical perspectives on MPB, we refer the reader to \cite{historia-mpb, faour-historia-musica-popular, cancao-tempo-1, cancao-tempo-2}.} MPB music typically exhibits a set of recurring characteristics:

\begin{itemize}
    \item Temporal arch: MPB's ``Golden-Era'' songs were written mostly from 1955 to the 1990s.

    \item Esthetics/character/mood: Frequently exploring social-political issues, especially during the military dictatorship (1964--1984), but also love stories, everyday facts, social and ecological causes, and many other themes.

    \item Harmony: Certainly the most salient of MPB's features, generally highly sophisticated, privileging complex, chromatic, dissonant, and dense chords, non-conventional progressions, occasionally modal (or modal/tonal) constructions, and modulations based on chromatic-mediant relations.

    \item Rhythm: Exploring mostly configurations of strongly syncopated Brazilian popular genres (\textit{samba}, \textit{choro}, \textit{baião}, \textit{bossa nova}, \textit{maracatu}, \textit{frevo}, etc.), sometimes merged with ``foreign''~ones (blues, rock, \textit{bolero}, \textit{tango} etc.).

    \item Melody: Employing complex relations between melody and harmony, especially by applying harmonic tensions at structural melodic points. 
\end{itemize}

However, MPB's universe can be more accurately defined by the works of some of its most well-known and distinguished composers, namely Antonio Carlos (Tom) Jobim (1927–1994), Ivan Lins (1945–), Chico Buarque de Hollanda (1944–), Edu Lobo (1943–), Caetano Veloso (1942–), Djavan (1949–), Milton Nascimento (1942–), João Bosco (1946–), Gilberto Gil (1942–), and Rita Lee (1947–2023), ten names that were selected to form the corpus of our project. Taken as a whole, these repertoires cover enormous diversity in every musical aspect one can imagine and span a very wide spectrum of personal styles. Despite this, we are convinced that it is possible to map a set of deeply grounded constructive features shared, to a greater or lesser extent, by many composers associated with MPB, forming a broad stylistic superset that links and contains individual compositional styles. The research question motivating the construction of the present dataset can be formulated as follows: \textit{which stylistic markers (related to rhythmic and melodic contour, harmony and the interaction between harmony and melody) function as shared norms within this musical aesthetic, and which ones diverge from these norms, thereby characterizing each composer's individual style?} We thus aim to define rigorously the contours of what we call the \textit{MPB Common Practice}, in analogy with Tymoczko's notion of the extended common practice \citep{tymoczko-geometry}. Almost paradoxically, despite its central role in twentieth-century Brazilian musical and cultural history, MPB has received surprisingly little structural and quantitative study. We propose this work as a first step toward closing that gap.

More precisely, the main goals of this paper are: 1) to briefly present the analytical models developed to encode, for our purposes, rhythmic, melodic, and harmonic information, as well as the melodic-harmonic relationship -- originally published in Brazilian Portuguese in \cite{almada-harmonia-jobim} and \cite{almada-melodia-jobim}, and presented here in shortened form in English for the first time, which is an additional contribution of this work; 2) to introduce the \textit{MPB Corpus}, the main contribution of this paper, a set of digital encodings of 500 compositions using those models; and 3) to present an exploratory data analysis illustrating its potential to help researchers better understand, from a musicological point of view, what MPB is. As for previous work, \cite{almada-harmonia-jobim} and \cite{almada-melodia-jobim} demonstrate the effectiveness of the models through an analysis of the works of Tom Jobim, and \cite{lamir-2024} presents a preliminary portion of the present dataset, restricted to melodic rhythm.

Beyond its musicological contribution, the MPB Corpus addresses a broader open-data need in MIR. Structural data for Brazilian popular music remain scarce, while copyright restrictions and the limited availability of reliable symbolic scores complicate the open dissemination of source materials. We therefore release the derived annotations in documented, machine-readable formats, with the relevant vocabularies and analysis code, enabling reuse and computational comparison without redistributing copyrighted scores or recordings.

The text is organized as follows. Section \ref{sec:corpus-design} presents the corpus design and methodology, including the selection of artists and related limitations; Section \ref{sec:analytical-models} presents the analytical models developed to encode rhythm, melody, harmony, and melody-harmony relationships, together with an encoding example; Section \ref{sec:organization-dataset} summarizes the organization of the corpus; Section \ref{sec:statistical-analysis} presents an exploratory data analysis across all analyzed domains, complemented by a permutation-based test of the observed rhythmic differences; Section \ref{sec:applications-dataset} proposes applications of the corpus; Section \ref{sec:corpus-development} discusses its development and long-term expansion; and conclusions are drawn in Section \ref{sec:conclusion}.

\section{Corpus design and methodology}
\label{sec:corpus-design}
As mentioned above, we selected ten composers to form the corpus, listed in chronological order of analysis: Tom Jobim, Ivan Lins, Chico Buarque, Edu Lobo, Caetano Veloso, Djavan, João Bosco, Milton Nascimento, Gilberto Gil, and Rita Lee. This selection reflects the authors’ perspective on the most significant composers in MPB, the consensus of Brazilian specialized music critics \citep{faour-historia-musica-popular, castro-a-noite-do-meu-bem}, and typical listeners' perspective. Two further reasons motivated restricting this stage of the investigation to these ten composers. First, they are among the most commercially successful representatives of the tradition, and many remain active today in live performance and new recordings. Second, and relatedly, high-quality published scores are widely available on the Brazilian editorial market, reliably supporting the analytical procedures employed here.

It should be noted that the selected group is predominantly male. This reflects in part the scope of the project, which focuses on compositional esthetics rather than performance practice. Consequently, several highly influential MPB artists (such as Gal Costa and Elis Regina) are not included, as their contributions are documented primarily through performance rather than through a body of published compositions. Moreover, although important women in MPB were also active as composers (for instance, Dona Ivone Lara, Maysa, and Joyce), reliable published scores for their works are often unavailable; including them would require prior transcription from audio recordings, which lies outside the methodological scope of this paper.

In the long term, the project aims to compile and analyze works by approximately fifty composers associated with the esthetics of MPB, which will improve the representation of women in our dataset. Expansion must nonetheless be left to future stages, for two main reasons. First, the analytical procedure is carried out almost entirely manually (see Sections \ref{sec:transcription-example} and \ref{sec:organization-dataset}), making data collection and analysis extremely time-consuming and labor-intensive even when reliable published scores are available. Second, for many important artists (such as Belchior, Jorge Ben Jor, and the female composers mentioned above), no reliable score sources are widely available, requiring transcription directly from audio recordings before analysis can take place.

The analysis follows specific guidelines: only the main melodic sections are considered, excluding instrumental introductions, interludes, and \textit{codas}; literal repetitions (such as \textit{da capo} instructions) are also ignored. This approach thus prioritizes what we term the \textit{nominal form} of compositions—their essential structural material—rather than their \textit{realized form}, which includes full arrangements. For the selection of works, we prioritized songbooks available on the Brazilian publishing market\endnote{ISBNs are provided exceptionally for these references, to identify the specific editions consulted.} \citep{songbook-milton, cancioneiro-jobim, songbook-djavan, cancioneiro-chico, songbook-lins, songbook-lobo, songbook-caetano, songbook-bosco, songbook-gil, songbook-rita}. Each score is carefully reviewed before transcription (considering form, notes, rhythms, and chords) so that errors are identified and corrected.

\section{Analytical models}
\label{sec:analytical-models}
We now present the models developed to encode melodic contour, melodic rhythm, harmony, and the relationship between harmony and melody. Introduced in \cite{almada-harmonia-jobim, almada-melodia-jobim} and successfully applied to the works of Tom Jobim, suggesting a potential to scale to a broader corpus within the MPB repertoire.

\subsection{Harmony}
\label{sec:am-harmony}
Our model for analyzing harmony divides into two branches, ``semantic'' and ``syntactic''. The ``semantic'' branch accounts for the structures of abstract \textit{chord types} (or CTs),\endnote{We distinguish chord types (abstract intervallic structures without any ``real'' root and pitch-class components) from specific (or ordinary) chords.} while the ``syntactic'' branch concerns the relationships between chords and chord types across multiple levels of organization. We now describe the model created for this research at the ``semantic''~level, the \textit{Genera of Chord Types}; the ``syntactic''~level is left for future work.

The \textit{Genera of Chord Types} (GCT)\endnote{A model with a homonymous acronym, General Chord Types, was proposed by \cite{cambouropoulos-gtc}, representing a chord by its root and the semitone intervals of its notes relative to that root. Our dataset does not employ this encoding, but we provide code in the supplementary material to convert our genealogical notation to it.} is based on 10 archetypal chord types (the basic triads and seventh chords), known as \textit{protochords}, each representing a set (or \textit{genus}) of generic chord types whose members include the protochord itself and all its possible variants. Derivation is governed by production rules and the recursive application of three transformations: ADD (\textit{addition} of a note to a chord type, such as a ``ninth'' to a major triad), SUB (\textit{substitution} -- e.g., replacing the ``third'' with the ``fourth'', yielding a ``dominant-with-suspended-4th'' chord type), and ALT (\textit{alteration} -- chromatic modification of a chord note, yielding, for example, a ``flattened thirteenth''). The 10 genera are named with the final letters of the Latin alphabet, to avoid confusion with the usual chord naming conventions; uppercase letters refer to major-mode protochords and lowercase letters to minor-mode ones, as shown in Table \ref{tbl-4-1}. For a more detailed description, see \cite{almada-harmonia-jobim}.

\begin{figure*}[!htb]
    \centering
    \includegraphics[width=0.85\textwidth]{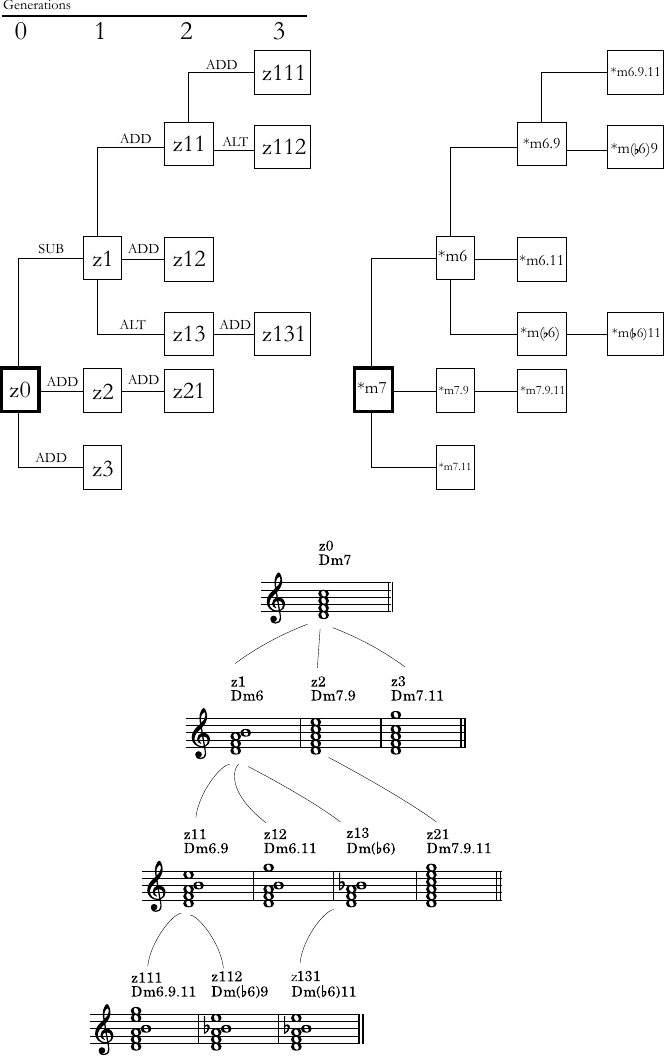}
    \caption{\textit{Genus} of protochord z0 depicted in three notations: genealogical (top left), chord-label (top right -- using the symbol * as a generic root.), and musical (below -- using ``D'' as the reference root).}
    \label{genealogia z}
\end{figure*}

\begin{table}[htb]
    \centering
    \begin{tabular}{@{}ccc@{}}
        \toprule
        \textbf{Genera} & \specialcell{Genealogical \\ notation} & \specialcell{Alphanumeric \\ notation} \\
        \midrule
        Z & Z0 & *M7 \\
        Y & Y0 & *7 \\
        X & X0 & *(\musFlat{}5)7 \\
        W & W0 & *(\musSharp{}5)7 \\
        V & V0 & * \\
        z & z0 & *m7 \\
        y & y0 & *$^\text{\o}$ \\
        x & x0 & *°7 \\
        w & w0 & *m(M7) \\
        v & v0 & *m \\
        \bottomrule
    \end{tabular}
    \caption{The 10 protochords of the GCT model, considering genealogical and alphanumeric notations. The symbol * stands for a generic root.}
    \label{tbl-4-1}
\end{table}

Protochords and variants of the same \textit{genus} are identified through two notations: (a) \textit{alphanumeric}, based on a roughly conventional notation among composers and instrumentalists,\endnote{The main feature of this notation is the ordering of numerical symbols in labels and the use of separating dots, in addition to parentheses for function note alterations: C6.M7.9, E4.7(\musFlat 9)13, F(\musSharp 5)7, etc. For more details, see \cite{almada-harmonia-jobim}.} and (b) \textit{genealogical}, a formula combining the \textit{genus} symbol with a numerical code identifying the lineage of the respective chord type. By convention, all protochords are numbered with a ``zero'' to the right of their letter-symbol, as shown in Table \ref{tbl-4-1}.

First-generation variant CTs are produced when transformational operations are applied to protochords. Consider, for example, protochord z0 (*m7 in alphanumeric notation, the symbol * representing a generic root) and its 10 CT variants, covering three generations, as illustrated in Figure \ref{genealogia z}. Under recursive application, first-generation variants in turn become parents of second-generation ones, and so on: CT z11 results from adding a ninth to z1 (giving *m6.9), and the third-generation z111 from adding an eleventh to z11 (giving *m6.9.11). To avoid redundancies and overpopulation, a set of constraints and rules limits the system to 161 variants, distributed across the \textit{genera} and generations as presented in Table \ref{tbl-4-2}. The main constraints are (for more details, see \cite{almada-harmonia-jobim}):

\begin{itemize}
    \item [1) ] The normative order of application of operations is SUB (if possible)--ADD--ALT;
    \item [2) ] Operation SUB is allowed only in three cases: *M7$\rightarrow$*6 (\textit{genus} Z), *m7$\rightarrow$*m6 (\textit{genus} z), and *7$\rightarrow$*4.7 (\textit{genus} Y);
    \item [3) ] An eleventh can be added only to CTs of \textit{genera} z, y, x, and w;
    \item [4) ] A thirteenth can be added only to CTs of \textit{genera} Y, W, X, and y;
    \item [5) ] A sharp-eleventh can be added only to CTs of \textit{genera} Z, Y, W, and V;
    \item [6) ] The application of ALT to tensions ``ninth'' and ``thirteenth'' is only possible in \textit{genera} Y, X, and W;
    \item [7) ] The application of ALT to the fifth is only possible in \textit{genera} Z and V.
\end{itemize}

As Table \ref{tbl-4-2} shows, these constraints leave some \textit{genera} quite sparse (such as ``w'', with only four members), while others are highly dense, notably the ``dominant'' genus ``Y'', with 59 elements -- explained by the combinatorial possibilities arising from its rich set of tensions and their alterations.

\begin{table}[htb]
    \centering
    \begin{tabular}{@{}cccccccc|c@{}}
        \toprule
        \textbf{Genera} & 0 & 1 & 2 & 3 & 4 & 5 & 6 & \textbf{Total} \\
        \midrule
        Z & 1 & 5 & 7 & 3 & - & - & - & 16 \\
        Y & 1 & 4 & 9 & 14 & 16 & 13 & 2 & 59 \\
        X & 1 & 3 & 5 & 6 & 4 & - & - & 19 \\
        W & 1 & 2 & 3 & 3 & 1 & - & - & 10 \\
        V & 1 & 4 & 5 & 2 & - & - & - & 12 \\
        z & 1 & 3 & 4 & 3 & - & - & - & 11 \\
        y & 1 & 3 & 3 & 1 & - & - & - & 8 \\
        x & 1 & 4 & 6 & 3 & 1 & - & - & 15 \\
        w & 1 & 2 & 1 & - & - & - & - & 4 \\
        v & 1 & 3 & 3 & 1 & - & - & - & 8 \\
        \midrule
        \textbf{Total} & 10 & 33 & 46 & 36 & 22 & 13 & 2 & 161 \\
        \bottomrule
    \end{tabular}
    \caption{Distribution of CTs across the 10 \textit{genera} (shown in the rows), considering a total of six possible generations (shown in the columns). The total number of variations for each \textit{genera} across all the generations is displayed in the last column; the total number of variations in each generation across all \textit{genera} is displayed in the last row.}
    \label{tbl-4-2}
\end{table}

The genealogical notation thus describes a variant's derivative position within its genus through a numerical sequence. The label Y2111 (related to CT *($\flat$9$.\sharp$11)13), for example, is a fourth-generation variant of genus Y, as the number of digits after the identifying letter indicates. Read from right to left, its position is: first variant (4th generation) of the first variant (3rd generation) of the first variant (2nd generation) of the second variant (1st generation) of Y0. For a complete list of CTs and their genealogical relations, see the supplementary material.

For each chord we also collected its alphanumeric notation and its functional analysis within the context of the piece. Functional categories follow a notation detailed in a lexicon, provided in the supplementary material.

The GCT framework also enables a more explicitly transformational perspective on harmony, grounded in the principles of \textit{Grundgestalt} and developing variation proposed by Schoenberg \citep{schoenberg-style, frisch-dev-var} and further elaborated in \cite{almada-mda}. Harmonic structure can then be interpreted in terms of transformations between abstract chord types rather than relations between surface-level chord tokens, allowing harmonic syntax to be studied at a higher level of abstraction: the relations between successive chord categories, and the transformations required to derive specific chords from more fundamental protochord structures.

This perspective also suggests compositional applications, since the transformations underlying a piece can be repurposed in generative settings to recreate stylistic nuances in a more abstract and controlled manner. Both this compositional direction and the syntactic analysis of the proposed corpus are ongoing lines of research within the project; preliminary results, focused on the works of Tom Jobim, can be found in \cite{almada-harmonia-jobim}.

\subsection{Pitch and rhythmic structures}
\label{sec:am-pitch-rhythm}
The \textit{Melodic Filtering Model} (MFM) is the component of our system responsible for melody, encompassing pitch and rhythmic structures. It rests on three principles: \textit{segmentation}, \textit{abstraction}, and \textit{encoding}. Segmentation assumes that understanding sequential information requires dividing it into smaller units, each with a certain degree of autonomy -- short musical ``phrases'' or, more neutrally, small groups in the sense of Lerdahl’s grouping structure \citep{lerdahl-gttm}; see also \cite{snyder-music-memory}. Such segments, analogous to Lerdahl’s groups, are referred to in the model as \textit{words}. Abstraction allows different structures to be grouped into equivalence classes with respect to a given comparison parameter. Encoding, finally, isolates pitch and rhythmic structures from the melodies analyzed, transforming them into abstract descriptions of these domains. 

\begin{figure}[htb]
    \centering
    \includegraphics[width=0.8\linewidth]{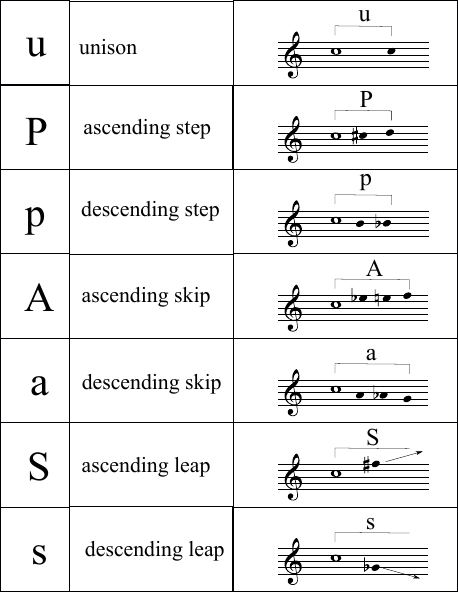}
    \caption{Alphabet of c-letters, their description, and an illustration of the distances encoded by each letter, starting from the C note.}
    \label{fig-celetters-tradu}
\end{figure}

For melody, each pair of contiguous pitches in a melodic line is encoded as sequences of seven \textit{contour gestures}, forming an alphabet of \textit{c-letters} (Figure \ref{fig-celetters-tradu}): note repetitions, stepwise motion (movement of one or two semitones), arpeggios (notes three to five semitones apart), and leaps (six semitones or more), both ascending and descending.\endnote{Our melodic contour modeling resembles Parsons Code \citep{parsons} but is more detailed, since ``up''~and ``down''~movements are further decomposed. The supplementary material provides code to convert our notation to Parsons Code (not reversibly) or to \cite{dowling} notation (``u''~= unison; ``t/T''~= step; ``s/S''~= skip; ``l/L''~= leap).}

\begin{figure}[!h]
    \centering
    \includegraphics[width=0.75\linewidth]{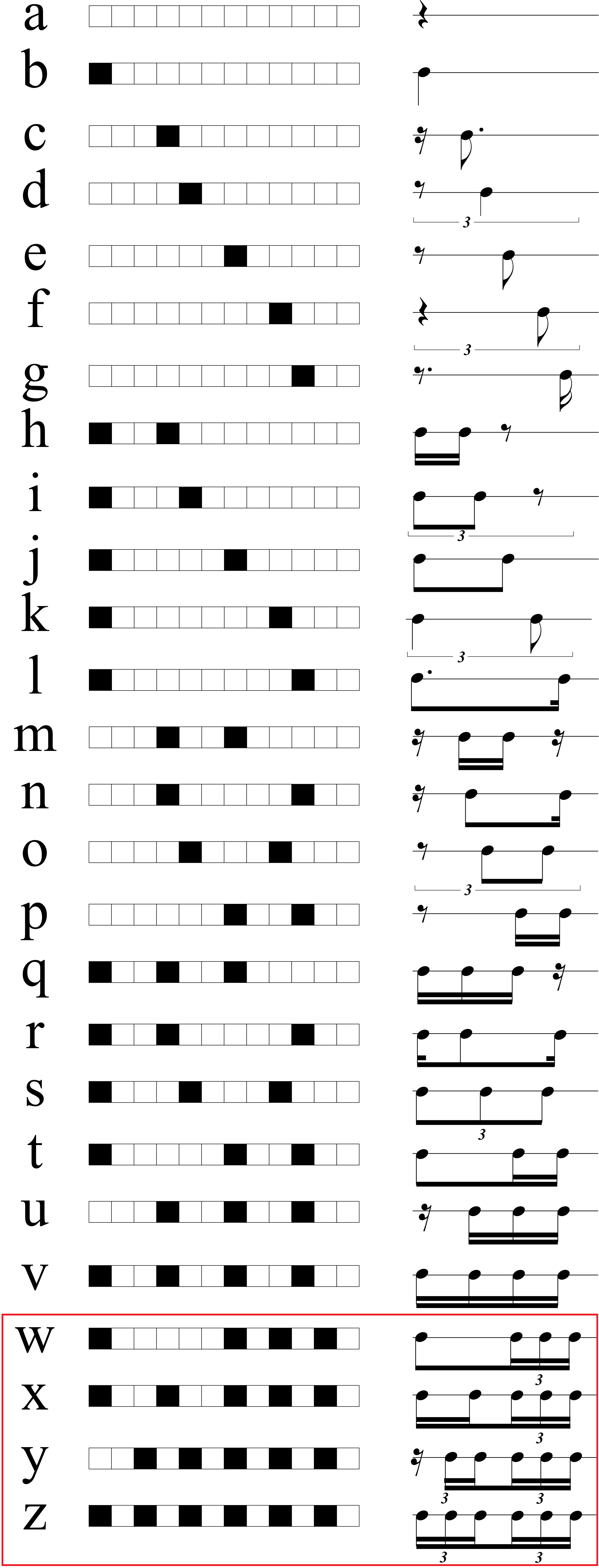}
    \caption{Alphabet of r-letters.}
    \label{fig-r-letters}
\end{figure}

The rhythmic model, briefly presented in \cite{lamir-2024}, is recalled and expanded here. The strategy subdivides a beat (quarter note) into 12 units and associates the possible configurations of IOIs (\textit{Inter-Onset Intervals}, the distances between attack points -- durations are not considered) occurring within this time window with standard units, called \textit{r-letters} in the model, forming an alphabet of 26 elements (Figure \ref{fig-r-letters}).

Note that the last four letters in Figure \ref{fig-r-letters} are highlighted. The r-letters from ``a'' to ``v'' proved sufficient to cover almost the entire repertoire of interest, but the MFM, like all frameworks presented in this paper, can also be used to analyze other corpora. An analyst may therefore rely on ``a'' through ``v'' and add as many ``wildcard letters'' as necessary to accommodate less common scenarios; the letters from ``w''~to ``z''~are four such suggestions.

Two important metrics are considered in relation to melodic contour and rhythmic structure, both yielding values between 0 and 1: the \textit{compensated intervallic economy index} (CIEI), which captures the profile of melodic contours within a c-word from the relative prevalence of more ``economical''~intervallic movements, and the \textit{countermetricity index} (CMI). \textit{Countermetric rhythms} are rhythmical patterns that deviate from or oppose the underlying metrical grid, often by emphasizing offbeats, generating syncopation, or creating tension against the regular pulse. The notion is particularly relevant for Afro-Brazilian genres, which highly influence MPB, where such rhythmic displacements are integral to the esthetic rather than exceptions; for a detailed discussion, see \cite{sandroni-feitiço}.

To compute the CIEI, each letter in a c-word $w$ is mapped to a numeric value: $0$ for unison (``u''), $\pm 1$ for ascending/descending step (``P''/``p''), $\pm 3$ for ascending/descending arpeggio (``A''/``a''), and $\pm 5$ for ascending/descending leap (``S''/``s''). The mean of these attributed values is then computed and its absolute value taken -- call this value $|\overline{d}(w)|$ -- and normalized to the interval $[0, 1]$ as:
\begin{equation}
    \mathrm{CIEI}(w) = 1 - \frac{|\overline{d}(w)|}{5}.
\end{equation}
The maximum is thus attained by a c-word that perfectly balances upward and downward motions (e.g., ``pPpP...'', ``aAaA...'', ``uuu...''), and the minimum by one consisting only of leaps with no balance at all (e.g., ``sss...'', ``SSS...'').  

For the CMI, analogously, a value is assigned to each letter of an r-word $w$ based on its rhythmical identity: ``n''~is the ``most countermetric''~r-letter, with an assigned value of one, and ``b''~the ``least countermetric'', with an assigned value of $0.1$. For the full list of values, see the supplementary material. Let $c(w)$ denote the sum of the corresponding values of the letters in $w$, and $\mathrm{len}^*(w)$ the length of $w$ minus the number of times the r-letter ``a''~appears within it. The countermetricity index is then given by
\begin{equation}
    \mathrm{CMI}(w) = \lambda\frac{c(w)}{\mathrm{len}^*(w)},
\end{equation}
where $\lambda = 0.8$ if $\mathrm{len}^*(w) > 1$ and the last letter of $w$ is ``b'', and $\lambda = 1$ otherwise. The intuition is that r-words rich in highly countermetric letters (``c'', ``e'', ``g'', ``m'', ``n'', ``p'', and ``u'') will have a high CMI, while an r-word ending with an onset at the beginning of the beat (letter ``b'') has its CMI reduced by 20\% due to the metrical strength of this ending.

\subsection{Relations between melody and harmony}
\label{sec:am-rel-mel-harm}
Our model establishes hierarchical relations between structural notes and the chords that harmonize them, according to the well-known chord/scale theory \cite{gonda}. The scale of a chord can be seen as a convenient repository of structural notes it may harmonize, which includes not only its arpeggio notes but also the so-called tensions and, eventually, substitute notes.

\begin{figure*}
\centering
\includegraphics[width=0.86\textwidth]{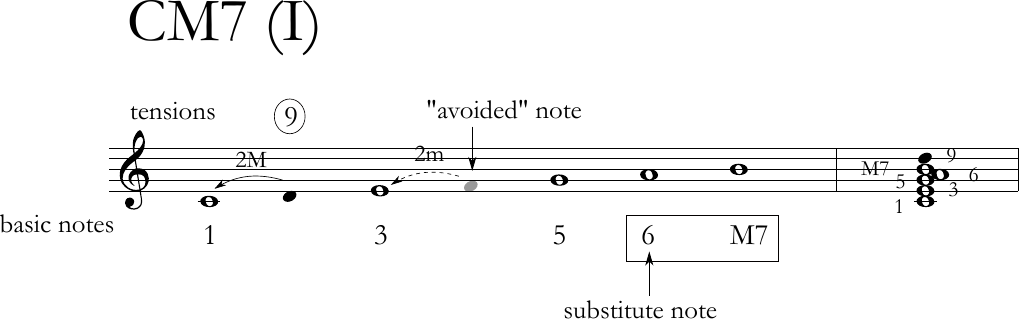}
\caption{Structure of the CM7 scale (I in C major) and identification of its structural notes.}
\label{fig-struct-notes-example}
\end{figure*}

Figure \ref{fig-struct-notes-example} illustrates this argument, considering the scale corresponding to the tonic chord (I) in C major, notated in its tetradic version as CM7.
On the left of the figure the notes are arranged in scale order. The white ones are the basic structural notes (identified below the staff), which form the arpeggio of the chord -- root (1), third (3), fifth (5), and major seventh (M7) -- including the sixth (6), which may eventually substitute the seventh. Tensions eligible to integrate the group of structural notes (above the staff) are evaluated by the practical rule of the ``two semitones'': a note not belonging to the basic arpeggio becomes a candidate harmonizable tension, and thus a structural note, if in the scale it lies \textit{two semitones above} the arpeggio note immediately to its left. If the distance is only one semitone, the note is excluded from the structural notes group and is generally called an ``avoided note''.\endnote{As with any rule, there are some exceptions. For their presentation and discussion, see \cite{almada-harmonia-jobim}.}

A verticalized representation of the scale, which highlights the hierarchy of structural notes, is shown to the right of Figure \ref{fig-struct-notes-example} (note F is omitted because it is not structural). From this set of structural notes established for the CM7 chord, functionally contextualized as the I degree, we can list all realizations of the chord that are equivalent in terms of functional position: C (triad), C.9 (triad with a ``ninth''~added), CM7 (the ``default configuration''), C6, CM7.9, and C6.9 (the last two incorporating the ``ninth''~tension). Notice that this equivalence is not necessarily compatible with the hierarchy present in the GCT model, whose goal differs from the one presented here.

From this perspective, we propose a broader understanding of the traditional notion of ``harmony'' (generally treated as an informal synonym of ``chords'') as the collection combining the notes that form a given chord with the structural notes present in the melody within that chordal context. Only structural notes are considered in this sum, disregarding any melodic inflections. We formally define an \textit{inflection} (or ``non-harmonic'' note) as a transitional note that, whether or not it belongs to the chord scale, is adjacent by a step to a clearly structural note that follows it, this structural note being called the \textit{target} of the inflection. Inflections are thus elements outside the chordal context and are not registered as structural. Beyond classifying notes as structural or inflections within a \textit{local} context (determined by the chords supporting the melodies), we also evaluate them according to the \textit{global} context (the current key), since melodies in popular music are predominantly diatonic and nondiatonic notes are, to a greater or lesser extent, digressions. This method normalizes melodies: the specific keys involved are not relevant, only the relationships between the notes and their respective diatonic scales.

\begin{figure}[htb]
\centering
\includegraphics[width=0.5\textwidth]{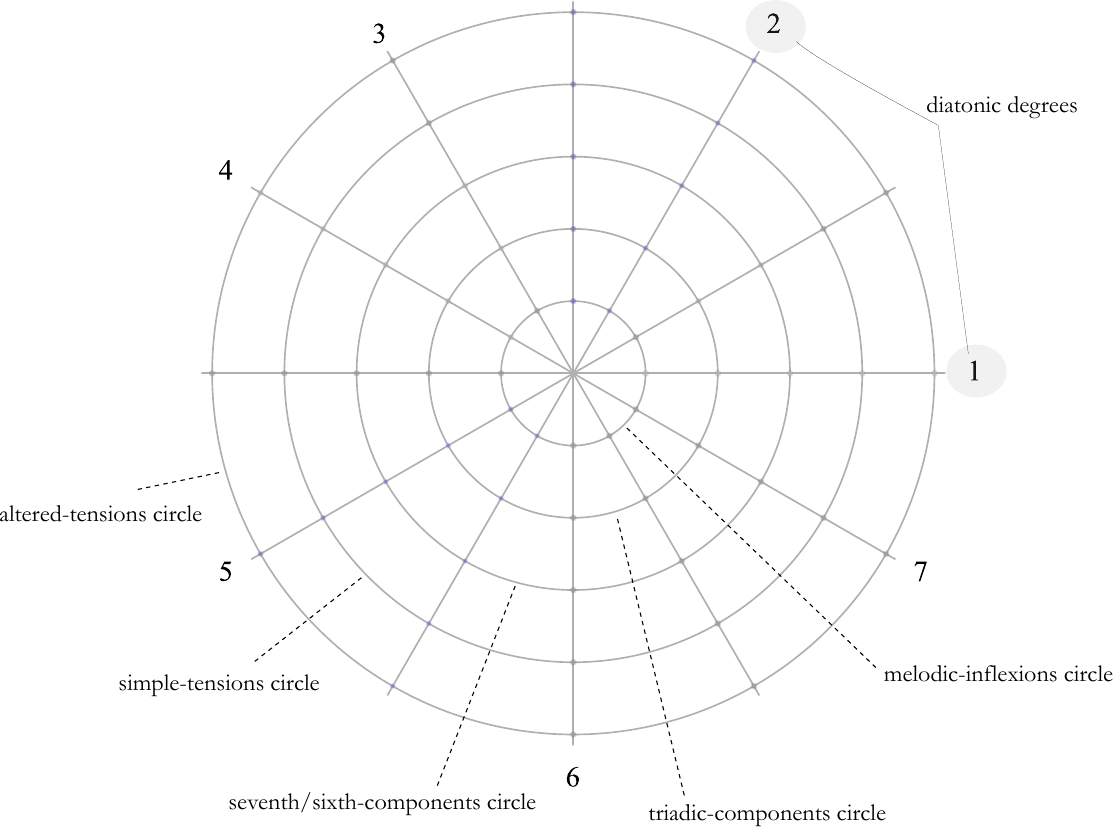}
\caption{Basic scheme of the NF web in major mode.}
\label{fig-nf-web}
\end{figure}

One graphical representation available within the model is presented in Figure \ref{fig-nf-web}. Named \emph{NF web} (after its shape, which vaguely resembles a spider's web; ``NF''~stands for ``note-function''), the graph consists of five concentric circles intersected by 12 rays arranged in angular increments of 30°, each ray corresponding to one of the notes of the chromatic scale. The scale degrees (in the figure, the major diatonic scale) represent the global context and are identified with Arabic numerals from 1 to 7, revealing their hierarchical superiority over the nondiatonic degrees on the intermediate axes. The concentric circles describe the local context: the smallest radius corresponds to the level of inflections (orbit 1), and the remaining circles follow the hierarchy of structural notes -- the basic notes forming the triad (orbit 2), the components of the seventh/sixth chord (orbit 3), the simple tensions (orbit 4), and finally the more loosely ``anchored'' altered tensions in the outermost orbit (5).

The NF web thus has 60 possible positions, the intersections of the circles and rays, denoting the possible combinations of local and global contexts. The color of a point at each intersection represents the proportion of the respective local-global combination in the piece being analyzed. It follows that melodies supported by more basic notes tend to occupy ``internal'' positions in the corresponding web, whereas melodies with more complex relationships between notes and chords (as, for example, in \textit{bossa nova}) tend to ``populate'' more ``external'' ones. 

An important metric for the NF model is the \textit{melodic anchoring index} (MAI), which quantifies, on average, the behavior of a melody in relation to its structural notes, taking as a parameter their degree of ``anchoring'' in the chords supporting them -- that is, how close or distant a given structural note is from its chordal context. The more basic the function performed by the note in the chord, the more ``anchored'' it will be. Only structural notes are taken into account in the calculation, disregarding the inflections present in the analyzed melody.

To compute the MAI, first recall the possible qualities that a note may have in a local context: ``\textit{root}'', ``\textit{third}'', and ``\textit{fifth}'', the basic triad function notes; ``\textit{tetradic}'', denoting the sixth or seventh degree of the underlying chord; ``\textit{simple tensions}'', the ninth, eleventh, thirteenth, or fourteenth degrees; and finally the ``\textit{altered tensions}'', encompassing the flat/sharp ninth, sharp eleventh, and flat thirteenth. Let $\mathcal{L}$ be the set of these six labels, and for each $\ell \in \mathcal{L}$ denote by $q_\ell$ the number of notes within that category in a given analyzed excerpt. The MAI is a weighted average of the $q_\ell$, each weight encoding the ``anchoring''~of the corresponding category:
\begin{align*}
    w_{\text{root}} &= 9\\
    w_{\text{third}} &= 7\\
    w_{\text{fifth}} &= 8\\
    w_{\text{tetradic}} &= 5\\
    w_{\text{simple tensions}} &= 3\\
    w_{\text{altered tensions}} &= 1.
\end{align*}
Denote by $w_m$ and $w_M$ the smallest and largest of these weights, respectively, and denote by $\overline{q}$ the average of $q_\ell$ weighted by $w_\ell$, for $\ell \in \mathcal{L}$:
\begin{align}
    \overline{q} = \frac{\sum_{\ell \in \mathcal{L}} w_\ell q_\ell}{\sum_{\ell \in \mathcal{L}} q_\ell - \eta},
\end{align}
where $\eta$ is the number of inflections within the analyzed excerpt. Finally, the MAI is a normalization of $\overline{q}$ within 0 and 1:
\begin{align}
    \text{MAI} = \frac{\overline{q} - w_m}{w_M - w_m}.
\end{align}
Values close to one indicate a predominance of structural notes within chords, whereas values close to zero indicate the opposite behavior. 

\subsection{An encoding example}
\label{sec:transcription-example}
We now summarize the model in action through the analysis of a short, hypothetical excerpt of the kind found in any piece of the MPB corpus (Figure \ref{ex-01_04}a). The analysis protocol follows these steps:

\begin{figure*}[!htb]
\centering
\includegraphics[width=0.81\textwidth]{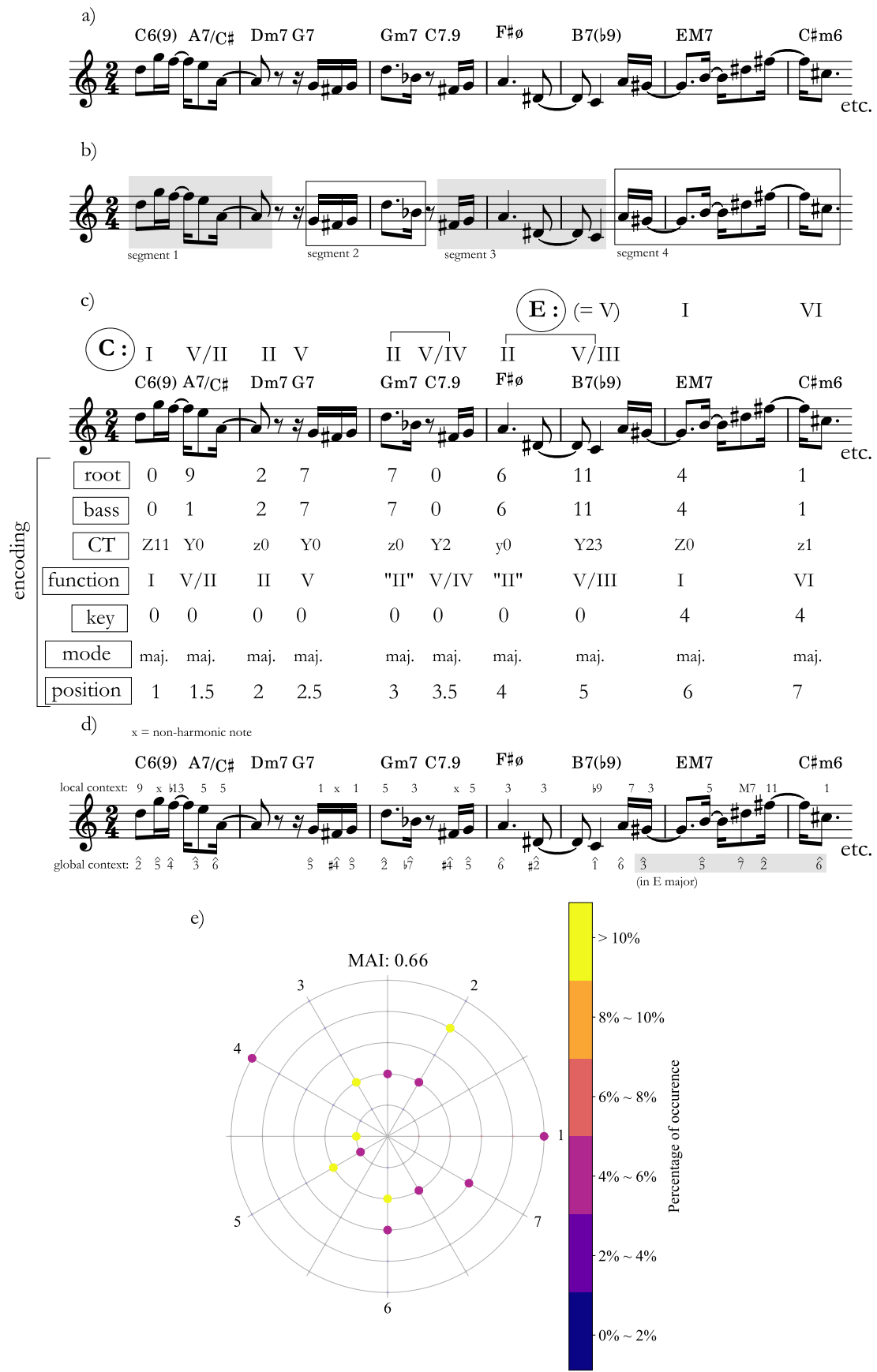}
\caption{Example of the analytical process: a) excerpt being analyzed; b) segmentation of the melody; c) harmonic analysis; d) relation between melody and harmony; e) NF web of this excerpt. For more details on the notation, see Section \ref{sec:analytical-models}.}
\label{ex-01_04}
\end{figure*}

\begin{enumerate}
    \item Subdivide the melodic line into logical and relatively autonomous segments (Figure \ref{ex-01_04}b).
    
    \item Encode pitch information inside each segment as c-words: $\langle$Apps$\rangle$ (segment 1), $\langle$pPSa$\rangle$ (segment 2), $\langle$PPsp$\rangle$ (segment 3), $\langle$pAAAa$\rangle$ (segment 4). Note that the transitions between segments are \textit{not} encoded.

    \item Encode rhythmic information inside each segment as r-words: $\langle$tn$\rangle$ (segment 1), $\langle$ul$\rangle$ (segment 2), $\langle$pbee$\rangle$ (segment 3), $\langle$pgnc$\rangle$ (segment 4). Recall that durations are immaterial in our model, only attack points are considered.
 
    \item Analyze the harmony of the piece, encoding for each chord (Figure \ref{ex-01_04}c): root and bass note (both as pitch classes), chord type (genealogical notation), function (Roman numerals), key (pitch class of the tonic), mode (major or minor), and metric position (bar numbers and fractions, where applicable).

    \item Analyze the functions that the notes perform in the harmonic structure, considering both the global (the key) and local (the chords that harmonize the notes) contexts (Figure \ref{ex-01_04}d -- functions in global and local context are displayed below and above the staff, respectively). In the local context, a note can be \textit{structural} (performing the function of root, third, fifth, sixth, seventh, ninth, eleventh or thirteenth) or non-harmonic -- an inflection, in our terminology -- performing no chordal function at all and represented by the symbol ``x''. The NF web related to this excerpt and its MAI are shown in Figure \ref{ex-01_04}e.
\end{enumerate}

In the case of an actual piece, the score, in its nominal form, is first transcribed into MIDI format using music notation software. The segmentation described in step 1) is implemented by inserting an identifying note into a register well above the melody (E6 in the present work); the presence of this note marks the end of a segment. Step 2) is then performed automatically by processing the corresponding MIDI file with a Python script. We attempted to automate step 3) as well, but particularities of the MIDI protocol made this unfeasible, so the encoding is performed manually; attempting the procedure with the MusicXML format instead is a possible direction for future work. Steps 4) and 5) are performed entirely by manual annotation. In particular, the harmonic analysis of each chord is the most critical component of the process and, to the best of our knowledge, cannot currently be reliably automated.

Throughout this process, occasional errors in source scores are identified and corrected during the analysis. Typical examples include the reinterpretation of chord symbols that, though notated so as to simplify harmonic execution (particularly on the guitar, a central instrument in MPB), represent a different chord in context; the incorporation of structurally relevant melodic notes into the underlying chord; and the correction of inaccurately notated harmonies. For examples of such corrections, see supplementary material.

\section{Organization of the MPB Corpus} \label{sec:organization-dataset}
The procedure illustrated in Section \ref{sec:transcription-example} was applied to 500 musical pieces evenly distributed among the following composers: Tom Jobim, Caetano Veloso, Edu Lobo, Chico Buarque, Milton Nascimento, Ivan Lins, Gilberto Gil, Djavan, Rita Lee, and João Bosco. To ensure methodological consistency across the dataset, particularly with respect to the harmonic analysis, the annotation was conducted by a single annotator, the first author of the present article, who has extensive expertise in music theory and harmonic analysis. After data collection, the dataset was systematically checked for inconsistencies such as mismatches between c- and r-words in the number of attack points, parameter values outside their valid ranges, and missing values; whenever such issues were identified, they were resolved by careful review of the original source material. This validation was assisted by the other two authors of this paper, who developed dedicated code to detect such inconsistencies automatically.

The dataset is archived on Zenodo \citep{mpb-corpus} and is hosted, together with its documentation, metadata, and accompanying code, on GitHub.\endnote{\url{https://github.com/ProjetoMPB/mpb-corpus}} We now present the structure of the main dataset files, which hold the raw output of the analytical procedure: tables of c-letters, r-letters, melody–harmony relationships, and harmony for each of the composers. In total, the dataset contains 8,426 c- and r-words, functional information for 17,053 chords, and melody–harmony relationships for 23,447 notes, making it the most detailed to date regarding Brazilian Popular Music.

For conciseness within the files that make up the dataset, composers are not referred to by their artistic name, but by shorter names, the correspondence being shown in Table \ref{tab:corpora-csv}. The database is organized into three \texttt{.csv} files, all of them containing the columns ``corpus\_id'', ``composition\_id'', and ``composition\_name'', referring respectively to the composer, identification number, and name of the composition under analysis. The repository provides these files in two equivalent layouts: \texttt{dataset\_aggregated/}, in which each file gathers all ten composers, and \texttt{dataset\_by\_corpus/}, in which the same data is split into one folder per composer. In all .\texttt{csv} files, each row represents an instance of the specific musical object being analyzed. We now detail the specific structure of each file corresponding to each musical parameter.

\begin{table}[htb]
\centering
\begin{tabular}{@{}ll@{}}
\toprule
corpus\_id & Composer's artistic name        \\ \midrule
JOBIM      & Tom Jobim         \\
LINS       & Ivan Lins         \\
CHICO      & Chico Buarque     \\
EDU        & Edu Lobo          \\
CAETANO    & Caetano Veloso    \\
DJAVAN     & Djavan            \\
BOSCO      & João Bosco        \\
MILTON     & Milton Nascimento \\
GIL        & Gilberto Gil      \\
RITA       & Rita Lee          \\ \bottomrule
\end{tabular}
\caption{Correspondence between the abbreviated corpus names (column ``corpus\_id'' within all the \texttt{.csv} files of the dataset) and the artistic names of the respective composers.}
\label{tab:corpora-csv}
\end{table}

For the melodic contours and melodic rhythm, the respective c- and r-words are contained in the file \texttt{contour\_rhythm.csv}, in columns ``c\_word''~and ``r\_word'', respectively. This file contains an additional column, ``word\_index'', representing the index of the respective segment within the analyzed piece.\endnote{A small number of occurrences of the r-word ``nan''~and of an empty c-word appear in the dataset. These entries should not be interpreted as missing data. The r-word ``nan''~represents a specific rhythmic pattern that occurs in four pieces: \textit{Acorda Amor} (Chico Buarque), \textit{Rebento} (Gilberto Gil), \textit{Bebel} (Tom Jobim), and \textit{Leva e Traz} (Ivan Lins). The empty c-word, in turn, represents a segment containing a single syllable, which also occurs in four pieces: \textit{Cálice} (Chico Buarque), \textit{Drão} (Gilberto Gil), \textit{Modinha} (Tom Jobim), and \textit{Os Povos} (Milton Nascimento). Therefore, when loading the files containing c-words and r-words, care must be taken not to interpret these entries as missing values.}

The relationship between harmony and melody is contained in the file \texttt{note\_function.csv}. %, its first lines being displayed in Table \ref{tab:melody-harmony-csv}.
Each line of the file regards one note, where columns ``scale\_degree'' and ``note\_function''~contain information about its relation with the global and local context of this note, respectively. Inflections are marked with an ``x''~within the column ``note\_function''. The column ``mode''~encodes the mode of the excerpt being analyzed (major/minor). For this musical parameter, only an excerpt of each piece was analyzed: more specifically, a contiguous segment of notes was examined in each piece and in only one mode (when mode changes occur). The rationale for this choice is that we empirically verified that such a sample was sufficient to capture the specific characteristics of each piece.

Finally, the harmonic analysis is contained in the file \texttt{harmony.csv}. Each line corresponds to one chord, columns ``root''~and ``bass''~contain the root and bass of the chord (in pitch-class notation), column ``chord\_type''~contains the genealogical notation of the chord, being translated in the usual alphanumeric notation in column ``chord\_symbol''. The column ``functional\_category''~explicits the function of the respective chord, columns ``key''~and ``mode''~contains the key (also in pitch-class notation) and mode of the excerpt analyzed. Finally, the position of the chord is described in column ``position'', where the number indicates the measure where the chord begins.

Alongside the analytical data, the repository provides a metadata file, \texttt{compositions.csv}, which links each composition to external music catalogues: a track on Spotify, the International Standard Recording Code (ISRC) of that recording, and, where available, the corresponding recording and work identifiers on MusicBrainz. These identifiers were obtained automatically and are offered as a convenience, allowing the reader to listen to the pieces and enabling the corpus to be cross-referenced with other datasets; they are not part of the analytical annotation.

\section{Exploratory data analysis} \label{sec:statistical-analysis}
We now present a non-extensive exploratory analysis of the proposed dataset. For each analyzed musical parameter, a series of tables and graphs are displayed, from which musicological hypotheses can be proposed; a permutation-based statistical test at the end of the section assesses the significance of the observed rhythmic differences. The purpose is to illustrate the potential of the dataset for both formulating and addressing important musicological questions about MPB, as well as to encourage further investigation in this direction. All tables and figures displayed here can be reproduced using the Python code in the supplementary material.

\subsection{Melodic rhythm} \label{sec:melodic-rhythm-stat}
\begin{figure*}[htb]
\centering
\includegraphics[width=0.95\textwidth]{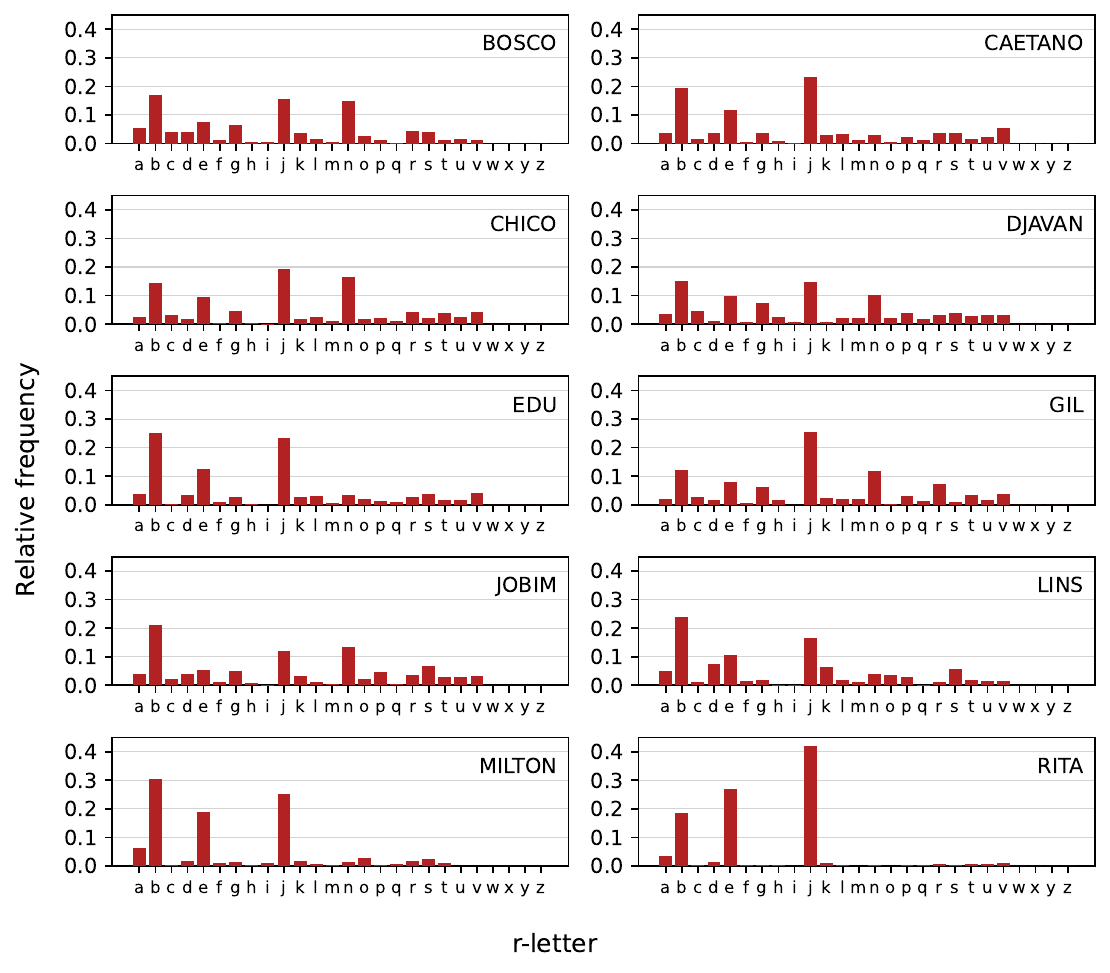}
\caption{Bar chart showing the proportion of each r-letter in each examined corpora.}
\label{fig:r-letters-dist}
\end{figure*}

The model for melodic rhythm was extensively presented in \cite{almada-melodia-jobim} and briefly explored on an excerpt of this dataset in \cite{lamir-2024}, whose discussion we now recall.\endnote{Minor numerical and graphical discrepancies with \cite{lamir-2024} may be found, because an extensive revision of the database corrected small analytical errors that do not affect the global statistics but are relevant for the more detailed analyses left as future work.} Figure \ref{fig:r-letters-dist} presents, for each corpus examined, the frequency of each r-letter. The data corroborates the practical observation that the r-letter ``b'' (characterized by a single attack akin to a ``rhythmic tonic degree'') predominates in nearly all repertoires. The presence of ``j'' is also notable in the works of Chico Buarque, Caetano Veloso, Milton Nascimento, Gilberto Gil, and Rita Lee, the last exhibiting the greatest percentage, nearly twice the average for this r-letter.% -- 42.6\% -- 

Information about r-letters can alternatively be conveyed using the \textit{metric profile}, the distribution of rhythmic attack points across a micrometric grid dividing each beat (taken here as a quarter-note unit) into 12 segments (cf. Figure \ref{fig-r-letters}); these data are shown in Table \ref{tab:perf-metrico} for the corpora studied. Fundamentally connected to rhythm, the metric profile is recognized as a crucial factor in highlighting the Brazilian ``lineage'' in MPB. Several observations stand out: (1) most repertoires predictably feature position 1 as the most frequent, with the exception of Rita Lee, where position 7 dominates; (2) the corpora of Jobim, Chico Buarque, Djavan, João Bosco, and Gilberto Gil closely align with each other, given their similar frequency of attacks at the first position (about $1/3$ of instances); (3) position 7, which divides the beat in half, sees significant activity in the Milton Nascimento corpus, considerably higher than the rest, and most notably in Rita Lee, marking the highest rate at 48.2\%; (4) the positions that segment the beat into three equal sections (5 and 9) are prominently represented in Tom Jobim and Ivan Lins (7.4\% and 7.5\%, and 10.7\% and 10.9\%, respectively), significantly more than in other repertoires, followed closely by João Bosco, with 6.7\% and 7.1\%.

\begin{table}
\centering
% \footnotesize
\begin{tabular}{@{}lllllll@{}} 
\toprule
   & 1             & 4    & 5    & 7             & 9    & 10   \\ \midrule
Bosco        & \textbf{31.5} & 17.0 & 6.7  & 18.1          & 7.1  & 20.2      \\ 
Caetano     & \textbf{37.6} & 10.9 & 4.7  & 27.9          & 4.5  & 14.4     \\ 
Chico      & \textbf{29.6} & 18.0 & 3.2  & 23.8          & 3.3  & 22.1      \\ 
Djavan            & \textbf{29.3} & 17.6 & 4.5  & 23.7          & 4.4  & 20.5       \\ 
Edu          & \textbf{40.7} & 8.3  & 5.5  & 27.6          & 5.5  & 12.4      \\ 
Gil      & \textbf{32.3} & 17.1 & 1.5  & 25.8          & 2.2  & 21.0    \\ 
Jobim         & \textbf{31.0} & 15.3 & 7.4  & 17.8          & 7.5  & 20.9     \\ 
Lins         & \textbf{38.2} & 6.7  & 10.7 & 23.2          & 10.9 & 10.4      \\ 
Milton & \textbf{46.5} & 3.5  & 5.9  & 33.4          & 5.8  & 4.9   \\ 
Rita          & 43.5          & 2.6  & 1.1  & \textbf{48.2} & 1.1  & 3.4     \\ \bottomrule
\end{tabular}
\caption{Metric profile by analyzed corpus, in percentage. The columns contain only the attack points present in the corpora. Percentages in bold correspond to the most prominent attack point on each corpus.}
\label{tab:perf-metrico}
\end{table}

\begin{table}
\centering
\footnotesize
\begin{tabular}{@{}llllllll|l@{}} 
\toprule
        & c   & e             & g   & m   & n             & p   & u   &  Total   \\ \midrule
Bosco   & 3.9 & 7.6           & 6.5 & 0.6 & \textbf{14.8} & 1.2 & 1.6 &  36.2    \\ 
Caetano & 1.6 & \textbf{11.7} & 3.8 & 1.3 & 2.8           & 2.1 & 2.1 &  25.4    \\ 
Chico   & 3.2 & 9.5           & 4.5 & 1.1 & \textbf{16.3} & 2.2 & 2.3 &  39.2    \\ 
Djavan  & 4.5 & 9.9           & 7.2 & 2.1 & \textbf{10.3} & 3.8 & 3.1 &  40.9    \\ 
Edu     & 0.2 & \textbf{12.6} & 2.8 & 0.6 & 3.4           & 1.1 & 1.5 &  22.3    \\ 
Gil     & 2.8 & 7.8           & 6.1 & 1.9 & \textbf{11.8} & 3.2 & 1.5 &  35.0    \\ 
Jobim   & 2.2 & 5.3           & 5.2 & 0.4 & \textbf{13.4} & 4.5 & 2.8 &  33.9    \\ 
Lins    & 1.0 & \textbf{10.8} & 1.8 & 1.1 & 3.9           & 2.8 & 1.3 &  22.8    \\ 
Milton  & 0.2 & \textbf{19.0} & 1.4 & 0.5 & 1.4           & 0.5 & 0.1 &  23.1    \\ 
Rita    & 0.3 & \textbf{26.9} & 0.5 & 0.1 & 0.5           & 0.5 & 0.8 &  29.6    \\ \bottomrule
\end{tabular}
\caption{Distribution of countermetric r-letters by corpus, in percentage. The last column shows the total occurrence of these r-letters for each composer. The most prominent countermetric r-letter in each corpus is highlighted in bold.}
\label{tab:contrametricas}
\end{table}

\begin{table}
\centering
% \footnotesize
\begin{tabular}{@{}lcccc@{}}
\toprule
        & \makecell{Word \\ length} & CIEI   & CMI     & MAI        \\ \midrule
Bosco   & 8                         & 0.856  & 0.559   & 0.704      \\
Caetano & 7                         & 0.873  & 0.463   & 0.721      \\
Chico   & 6                         & 0.887  & 0.564   & 0.704      \\
Djavan  & 6                         & 0.863  & 0.558   & 0.643      \\
Edu     & 8                         & 0.867  & 0.463   & 0.623      \\
Gil     & 8                         & 0.880  & 0.552   & 0.714      \\
Jobim   & 6                         & 0.888  & 0.531   & 0.582      \\
Lins    & 7                         & 0.867  & 0.446   & 0.649      \\
Milton  & 6                         & 0.898  & 0.388   & 0.684      \\
Rita    & 5                         & 0.886  & 0.429   & 0.741      \\ \bottomrule
\end{tabular}          
\caption{Statistics related to melodic contour, melodic rhythm, and the relationship between melody and harmony. The columns represent, respectively: length of the most common c- and r-word, CIEI, CMI, and MAI indexes.}
\label{tab:c-statistics}
\end{table}

\begin{figure*}[htb]
\centering
\includegraphics[width=0.95\textwidth]{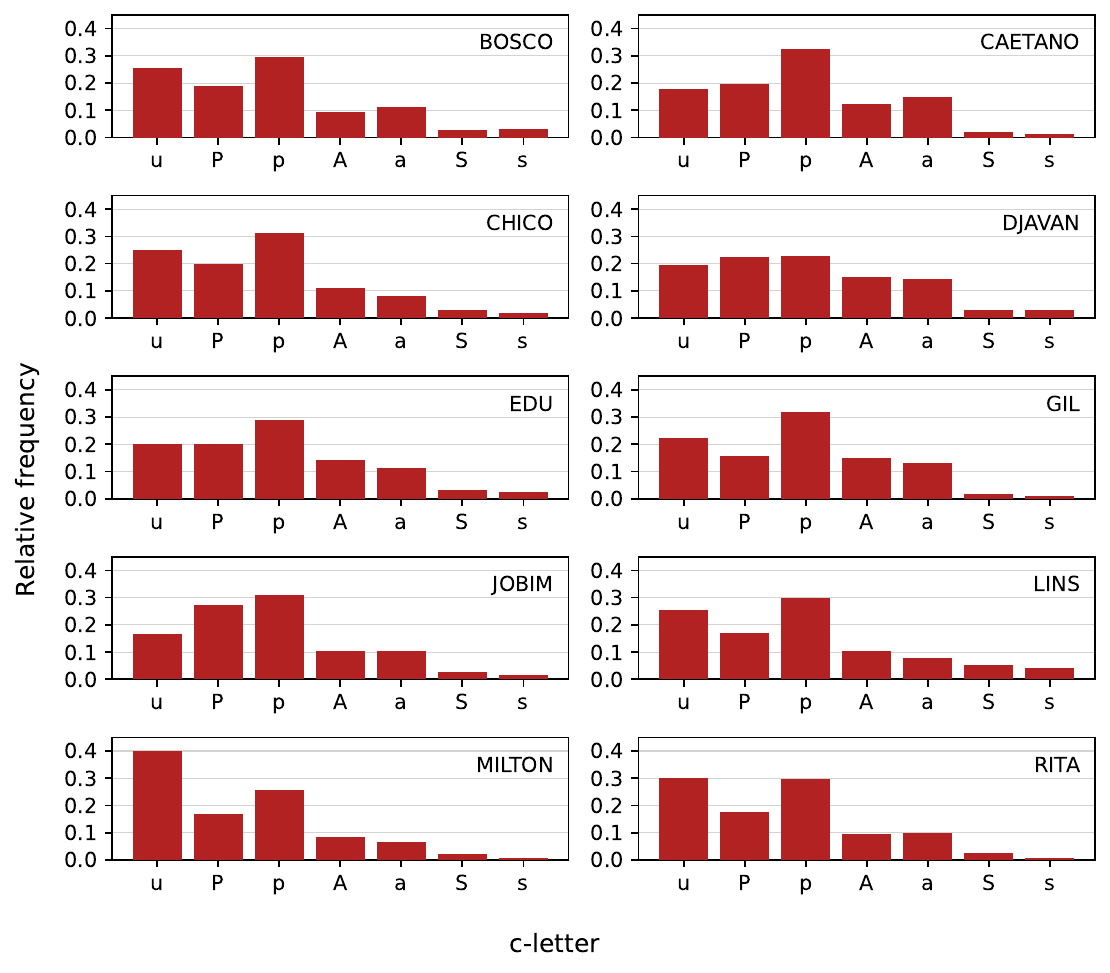}
\caption{Bar chart showing the proportion of each c-letter in each examined corpora.}
\label{fig:c-letters-dist}
\end{figure*}

Table \ref{tab:contrametricas} outlines the concurrence of countermetric r-letters. The emphasis on the countermetric r-letter ``e'' suggests a jazz influence in the rhythmic structure of Ivan Lins and Edu Lobo. Milton Nascimento also uses this r-letter frequently, which stems not only from jazz (a strong influence) but possibly from folk music, sacred music, or even Beatles songs; for Caetano Veloso, and notably Rita Lee, one might instead argue for rock and rhythm-and-blues. The collections of Jobim, Chico Buarque, João Bosco and, to a slightly lesser degree, Djavan and Gilberto Gil clearly indicate a closer relation to \textit{samba}, represented by the dominance of the r-letter ``n''. Overall, the substantial use of countermetric r-letters (around one-third of all occurrences) signals a high level of rhythmic syncopation, consistent with the CMI index displayed in Table \ref{tab:c-statistics}: for Chico, Bosco, Djavan, Gil and Jobim it is higher than $0.53$, probably due to the strong presence of \textit{sambas} in these repertoires.

\subsection{Melodic contour}
\label{sec:melodic-contour}
Figure \ref{fig:c-letters-dist} shows the frequency of each c-letter within each of the corpora examined. Although all repertoires converge toward ``stepwise motion''~as the preferred type of melodic motion, suggesting a shared stylistic rule, it is intriguing that in the corpora of João Bosco, Caetano Veloso, Chico Buarque, Djavan, Edu Lobo, Gilberto Gil, Tom Jobim, and Ivan Lins the most frequent c-letter is ``p''~(descending step). Whether the predominance of descending stepwise motion characterizes Brazilian popular music more broadly is an interesting avenue for further exploration. In the Milton Nascimento corpus, by contrast, the predominant melodic motion by a wide margin is note repetition (``u''), reflecting the notable stasis that prevails in the composer's melodic construction and is a defining trait. The Rita Lee repertoire is similar, differing only in that the percentage of ``u''~is lower, closely followed by the c-letter ``p''.%~(29.6\%).

\begin{figure*}[htb]
\centering
\includegraphics[width=0.95\textwidth]{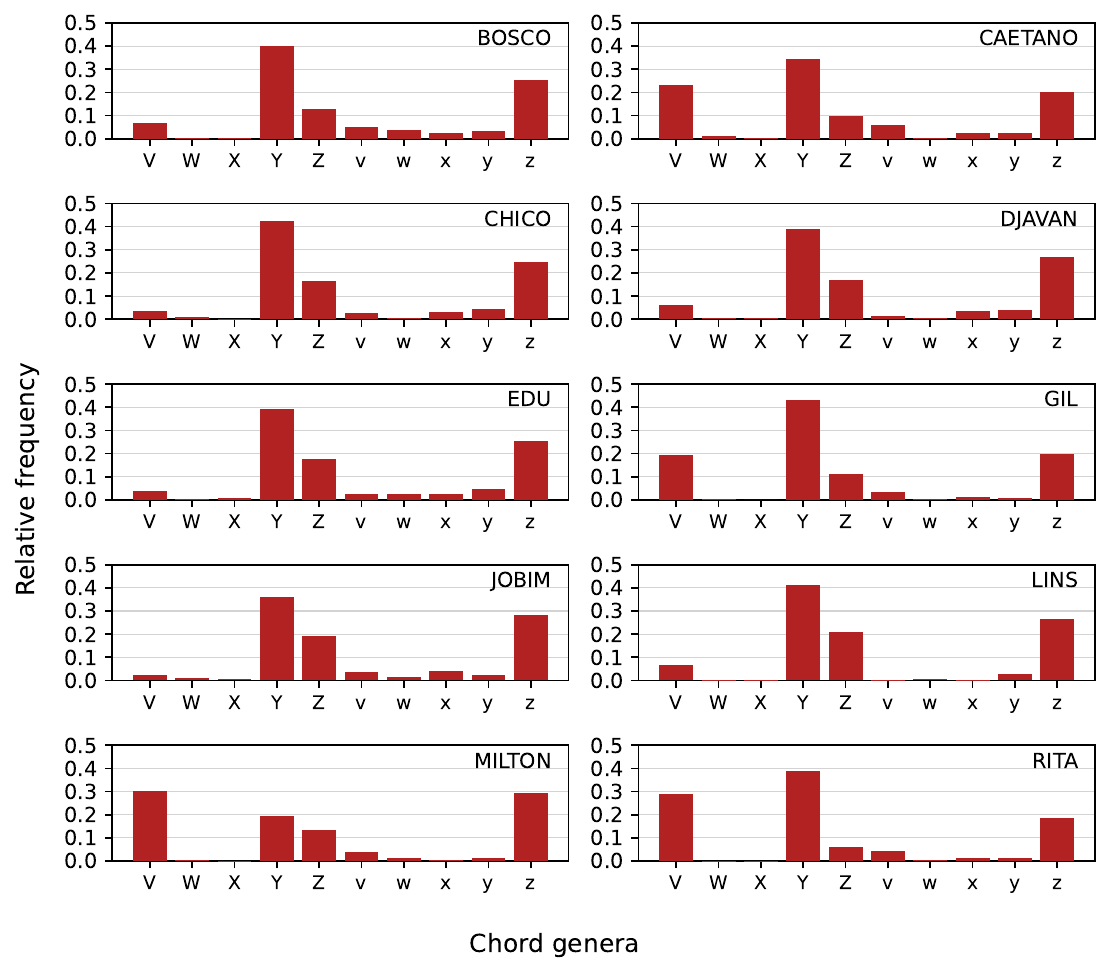}
\caption{Bar chart showing the proportion of each chord \textit{genera} in each examined corpora.}
\label{fig:chord-type-dist}
\end{figure*}

The compensated intervallic economy index (CIEI) measures, in essence, the degree of melodic stability. The values obtained, shown in Table \ref{tab:c-statistics}, broadly support the general notion that a kind of \textit{compensatory law} may be at work in melodic lines, which would ideally tend toward rectification, with leaps and dips cancelling each other out: all repertoires exhibit relatively high and very similar indices. The most common number of attack points in each melodic segment, also shown in Table \ref{tab:c-statistics}, converges toward six or seven melodic articulations, in agreement with previous studies on melody segmentation \cite{snyder-music-memory}.

\subsection{Harmony}
Figure \ref{fig:chord-type-dist} illustrates the relative frequency of the ten \textit{genera} of chord types in all corpora analyzed. An initial interesting finding is the convergence of almost all corpora on the prominence of \textit{genus} Y, which undoubtedly reflects shared tonal syntactic properties and, more specifically, the presence of primary and secondary dominant preparations in these repertoires. \textit{Genus} V is strongly present in Caetano Veloso, Gilberto Gil, and Rita Lee, consistent with their greater adherence to the pop-rock esthetic compared to the other MPB composers; in Milton Nascimento's \textit{corpus}, which also presents a considerable proportion of triads, this is probably due to the predominance of modal syntax in the songs.

\begin{figure*}[htb]
\centering
\includegraphics[width=0.90\linewidth]{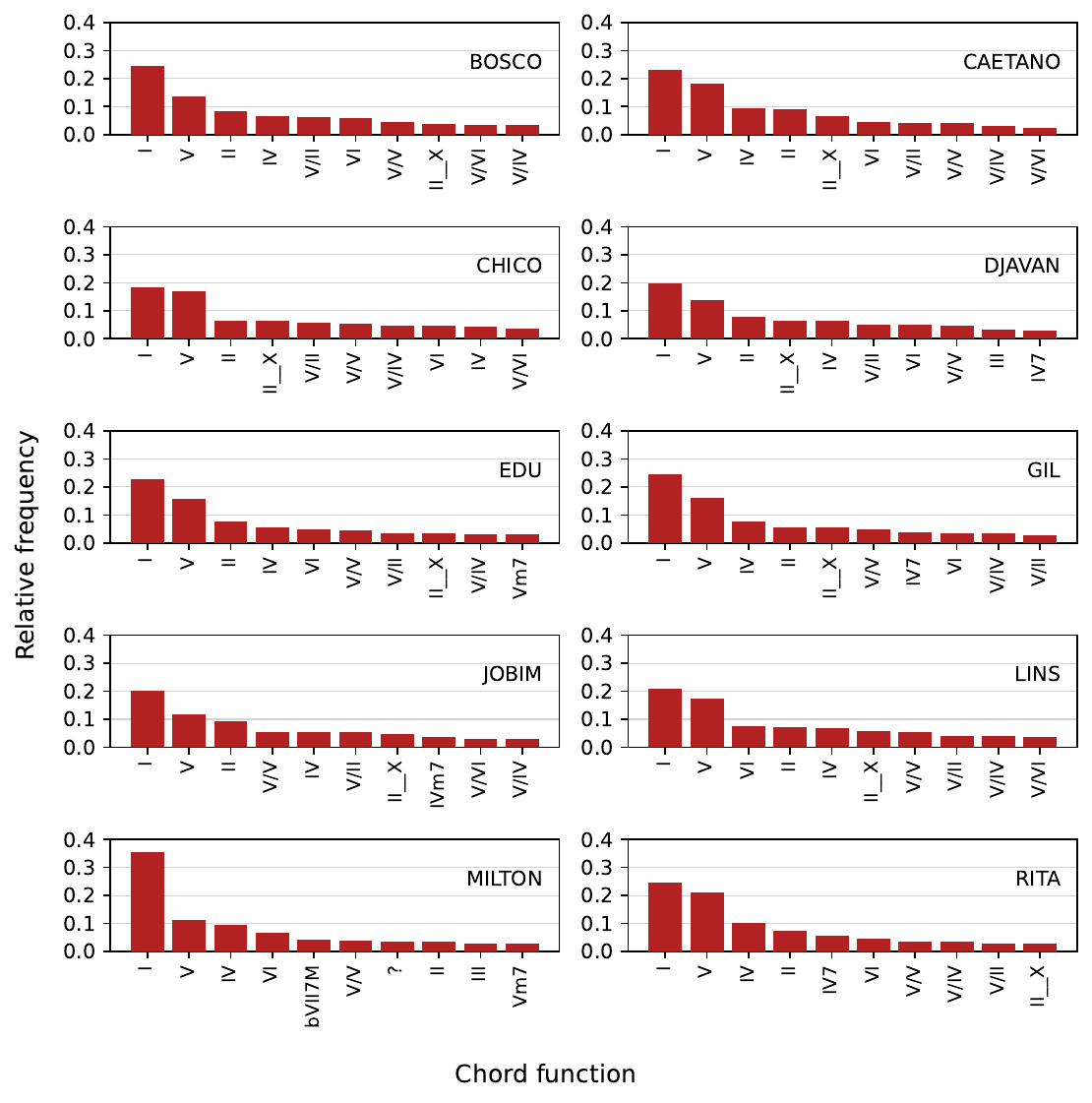}
\caption{Bar chart showing the proportion of the most common chord functions in each examined corpora.}
\label{fig:most-common-functions-dist}
\end{figure*}

Figure \ref{fig:most-common-functions-dist} illustrates the ten most common functional categories in all corpora analyzed. Not surprisingly, I and V (in this order) are significantly more common than the other categories in every corpus; IV, II, and VI (this last only in Ivan Lins) share the third position of the rankings, followed by different options, especially secondary dominants. This suggests that the basic functions, performed by central diatonic pillars, solidly ground the ``harmonic mass''~of MPB music shared by all the composers, while rarer chromatic, eccentric chords -- particular ``spices'' -- can be associated with individual stylistic preferences, a point to be properly examined in the continuation of the research.

\subsection{Relationship between harmony and melody}
\begin{figure*}[htb]
\centering
\includegraphics[width=0.90\linewidth]{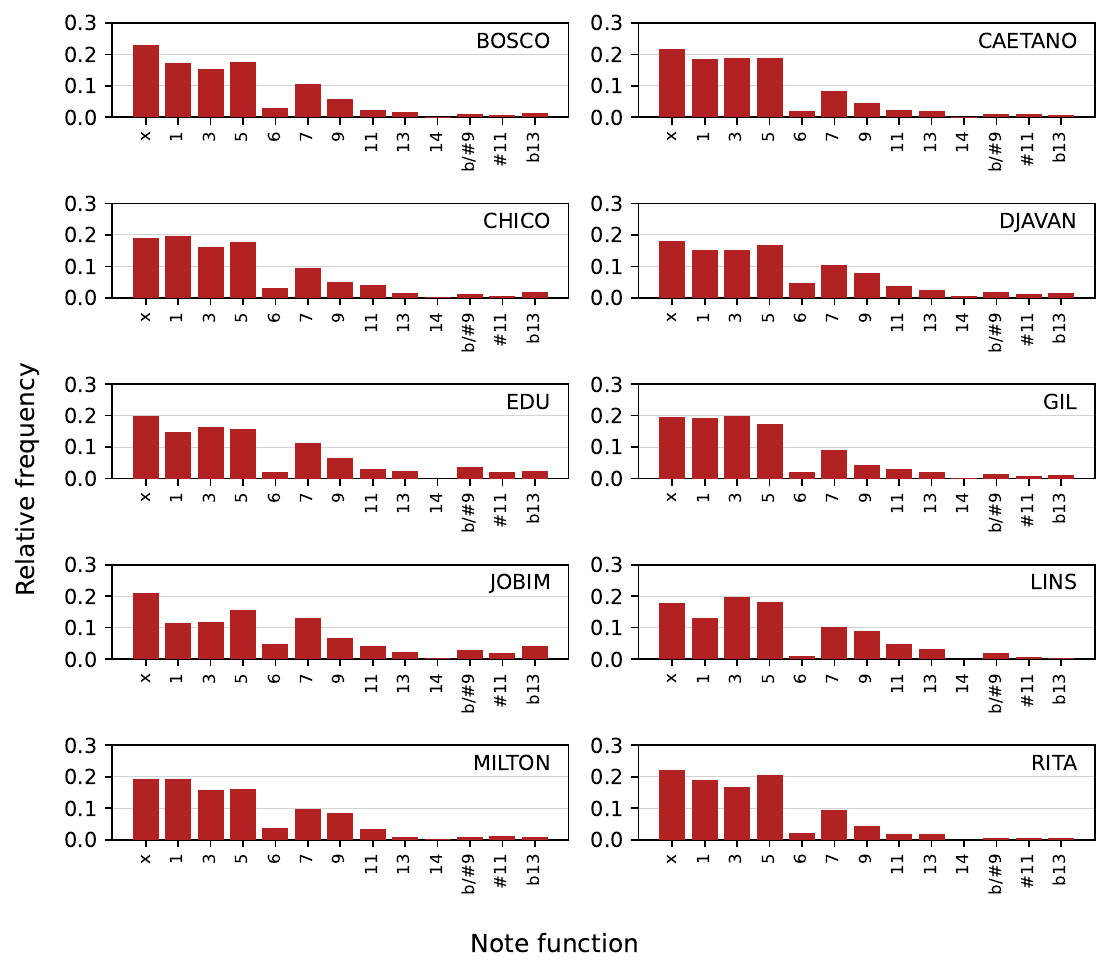}
\caption{Bar chart showing the proportion of each note function in each examined corpora.}
\label{fig:note-function-dist}
\end{figure*}

The relative frequency of each note-function in the corpora analyzed is shown in Figure \ref{fig:note-function-dist}. We first observe a certain uniformity in the distribution of inflections, which correspond on average to one-fifth of the melodic notes. Considering the classes of structural notes separately: (a) Triads (1, 3, 5) exhibit a similar distribution across all corpora, with some apparent exceptions such as Djavan, Edu Lobo, Jobim, and Ivan Lins, whose confirmation through statistical tests is left for future work; (b) Tetrads (6, 7) show a relative homogeneity among all repertoires, probably due to the frequency of the chord \textit{genus} Y noted above in the evaluation of harmonic attributes; (c) Simple tensions (9, 11, 13, 14) are likewise homogeneous across the repertoires, with the exception of Djavan, Ivan Lins, and Tom Jobim, which balances, in a way, their lower use of triadic notes; (d) Altered tensions (\musFlat{}9/\musSharp{}9, \musSharp{}11, \musFlat{}13):\endnote{In our analysis, the altered tensions \musFlat{}9 and \musSharp{}9 are grouped into a single category.} unlike the simple-tension class, this one is less distinctive, showing relatively similar distributions across all repertoires, with the exception of Edu Lobo and Tom Jobim, again compensating for their lower use of triadic notes.

\begin{figure*}[htb]
\centering
\includegraphics[width=0.9\textwidth]{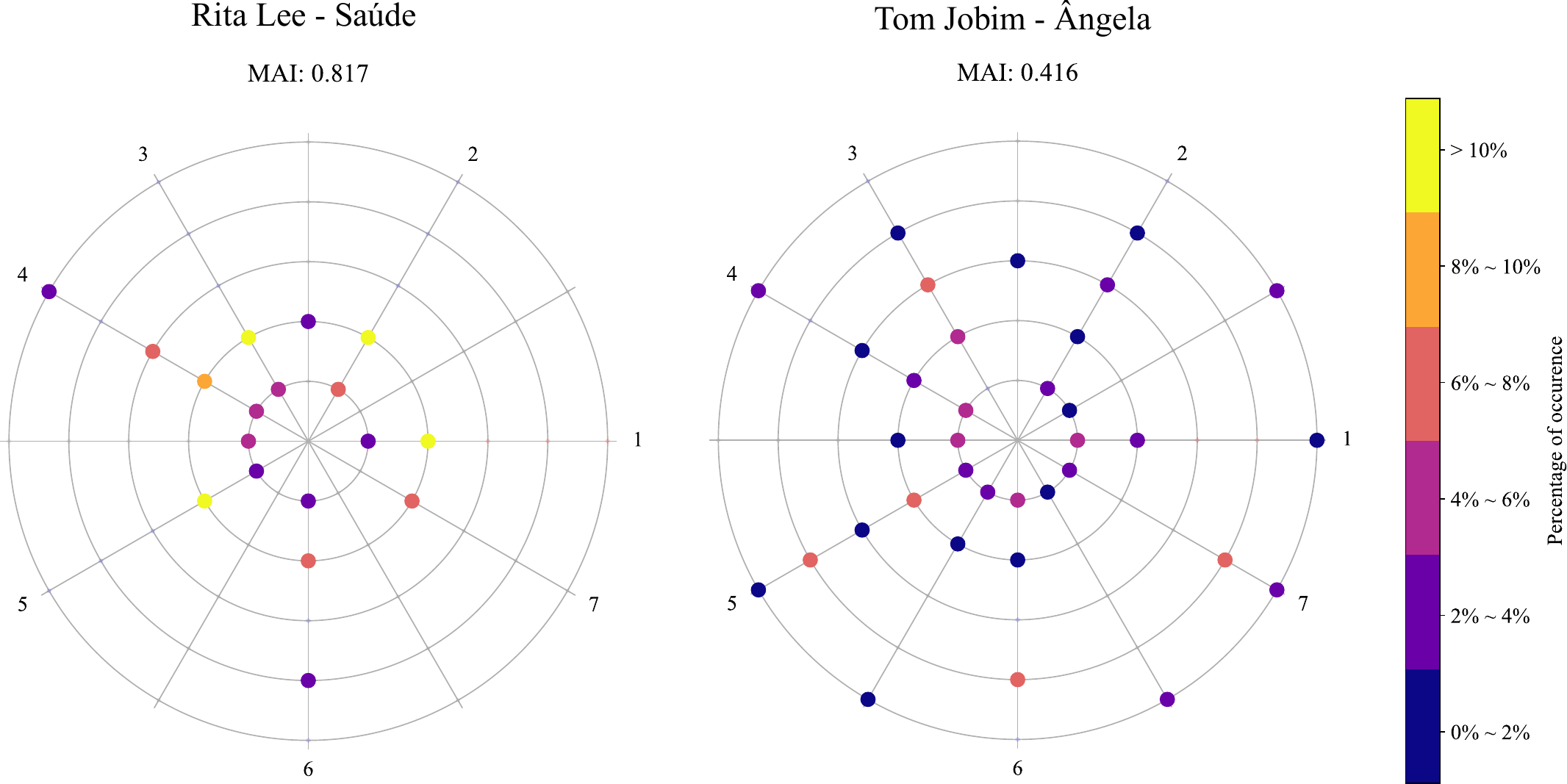}
\caption{NF web of two pieces: \textit{Saúde}, by Rita Lee (left panel), and \textit{Angela}, by Tom Jobim (right panel). Their respective MAI are displayed above the webs.}
\label{fig-teia-rita-jobim}
\end{figure*}

In summary, the corpora of Djavan, Edu Lobo, Ivan Lins, and Tom Jobim (especially) display a higher percentage of tensions than the other repertoires, suggesting that they may form a special subgroup characterized by more complex melody–harmony relationships. We expect to examine this hypothesis more rigorously from a clustering analysis perspective in future work.

Figure \ref{fig-teia-rita-jobim} depicts the NF webs of two songs with very distinct MAI values: Rita Lee's \textit{Saúde} (0.817) and Jobim's \textit{Angela} (0.416). The two graphs make the reason for such a discrepancy clear: in the former, most occurrences of NFs are distributed on the inner circles, denoting simpler relations between notes and chords, whereas Jobim's piece occupies a larger number of positions, especially in the more external circles, revealing more varied, eccentric, and complex relations.

\subsection{Statistical assessment of rhythmic profiles}
To illustrate one possible application of the proposed dataset, we present a brief statistical assessment of the rhythmic profiles of the selected composers. The purpose is not to provide a comprehensive musicological investigation (this is left for future work) but rather to demonstrate how the dataset can be employed to quantitatively compare stylistic characteristics across composers. Rhythm was chosen as a representative example because it is widely recognized as one of the defining features of Brazilian popular music, playing a central role in the characterization of genres, compositional styles, and individual musical identities, and because the visualizations presented above suggest noticeable differences in the distributions of rhythmic patterns across composers, motivating the question of whether these differences are statistically significant. Similar analyses may be carried out using the remaining musical descriptors available in the corpus.

Let $d(x_i,x_j)$ denote the Jensen-Shannon distance \citep{lin-jensen-shannon} between the frequency distributions of a musical descriptor of two musical pieces, given by
\begin{equation}
    \sqrt{\frac{1}{2}\left[D_{KL}\left(x_i, \frac{x_i + x_j}{2}\right) + D_{KL}\left(x_j, \frac{x_i + x_j}{2}\right)\right]},
\end{equation}
where $D_{KL}$ denotes the Kullback-Leibler divergence between two probability distributions, and $x_i$ and $x_j$ denote the normalized frequency distributions of the musical descriptor for musical pieces $i$ and $j$, respectively. To quantify the degree of separation between composers, we define the statistic
\begin{equation}
    R = \frac{\displaystyle \frac{1}{N_B} \sum_{\substack{i<j, c_i\neq c_j}} d(x_i,x_j)}{\displaystyle \frac{1}{N_W} \sum_{\substack{i<j, c_i=c_j}} d(x_i,x_j)},
\end{equation}
where $c_i$ and $c_j$ are the corresponding composers of pieces $i$ and $j$, and $N_W$  and $N_B$ are the numbers of within-composer and between-composer pairs of musical pieces, respectively.

The rationale behind this statistic is straightforward. If the distribution of the employed feature contains little stylistic information, the average distance between pieces by different composers should be comparable to that between pieces by the same composer, yielding values of $R$ close to one. Conversely, if composers exhibit distinctive distributions of the chosen feature, pieces by the same composer are expected to be more similar to one another than to pieces by different composers, resulting in $R >1$. Therefore, $R$ provides a simple and interpretable measure of the extent to which feature distributions discriminate between composers, while naturally accounting for the variability observed within each composer's repertoire.

To assess whether the observed value of $R$ within the dataset, denoted by $R_{obs}$, could reasonably arise by chance, a permutation test can be performed \citep{good-perm-test}. Under the null hypothesis that the chosen musical feature is unrelated to the composer's identity, the composer labels assigned to the musical pieces are exchangeable, so samples from the null distribution for $R$ can be generated by repeatedly permuting the labels at random, preserving the number of pieces associated with each composer, and recalculating the statistic after each permutation. Comparing $R_{obs}$ with the resulting empirical null distribution then yields the $p$-value, estimated as the proportion of permuted statistics greater than or equal to the observed value. More specifically, the Monte Carlo estimate is given by
\begin{equation}
    p = \frac{1 + \#\{R_{\mathrm{perm}} \ge R_{\mathrm{obs}}\}}{B+1},
\end{equation}
where $B$ denotes the number of permutations and $R_{\mathrm{perm}}$ the statistic obtained under each random permutation. This estimator avoids zero $p$-values and provides an unbiased Monte Carlo estimate of the exact permutation $p$-value \citep{good-perm-test}.

\begin{figure}[htb]
\centering
\includegraphics[width=0.95\linewidth]{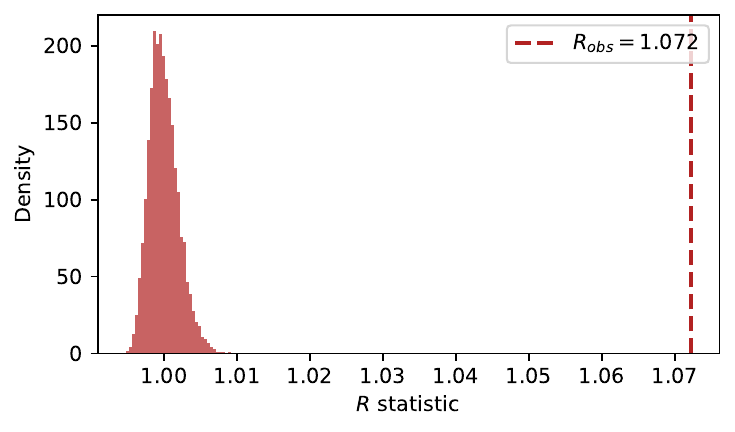}
\caption{Empirical null distribution of the statistic $R$ obtained from $10,000$ permutation samples, shown as a histogram. The red vertical line indicates the observed value $R_{obs}$.}
\label{fig:r-dist-rhythm}
\end{figure}

For this analysis, the selected musical feature was the distribution of r-letters across musical pieces. The observed statistic was $R_{obs} = 1.072$, and none of the $B = 10,\!000$ permuted statistics exceeded it; Figure \ref{fig:r-dist-rhythm} shows the empirical null distribution of $R$ together with the observed value. The estimated permutation $p$-value was smaller than $10^{-4}$, so the null hypothesis that r-letter distributions are independent of composer identity can be rejected at the adopted simulation resolution: the observed differences between composers are unlikely to have arisen by chance alone. These results indicate that r-letter distributions carry statistically significant stylistic information and illustrate the potential of the proposed dataset for quantitative studies of composer similarity and discrimination.

Extending this analysis to other musical descriptors, and relating the resulting patterns to clustering and other exploratory techniques, is an important direction for future work, with the potential to yield further musicological insight into stylistic organization in Brazilian popular music.

\section{Potential applications of the MPB Corpus}
\label{sec:applications-dataset}
The proposed dataset and the analytical framework developed to encode musical information are innovative and encompass aspects rarely discussed from a computational musicology perspective. We now highlight some of the musicological, analytical, and computational applications that both contributions enable.

From a musicological perspective, the four layers described by the analytical models (melodic contour, melodic rhythm, harmony, and melody-harmony relationships) allow the construction of comparative datasets across Brazilian musical genres (\textit{samba}, \textit{bossa nova}, \textit{baião}, \textit{frevo}, among others), supporting systematic investigation of the stylistic relations between them. Combined with metadata such as the year of release, these representations also facilitate diachronic studies of the historical evolution of MPB’s musical language, providing a quantitative basis for examining how stylistic traits changed across decades.

Another promising usage of the dataset concerns the relationships among contiguous musical elements, which play a fundamental role in musical structure (for example, functional chaining within harmonic progressions) but are not directly addressed in the present encoding. Such sequential relationships correspond to what may be understood as the \textit{syntactic level} of the dataset. These patterns can be explored using probabilistic models such as Markov models or Hidden Markov Models, enabling the analysis of transition matrices involving r-letters, c-letters, functional categories, note-functions, or other relevant musical units. Investigating these transitions may provide valuable insight into the compositional processes of individual composers, as well as the structural principles shared across the MPB repertoire.

In the context of MIR, the MPB Corpus offers structured symbolic data suitable for developing musical segmentation models that jointly exploit melodic, rhythmic, and harmonic dimensions, and for training classifiers for genre and subgenre prediction.

The corpus also has significant pedagogical value. In courses in computational musicology, it can serve as a concrete example of multilevel encoding, allowing students to explore symbolic representations that operate at complementary analytical scales. Its organization also makes it useful as a collection of analytical examples for courses in harmony, ear training, and stylistic analysis focused on MPB, enriching teaching materials with rigorously encoded repertory.

Beyond these analytical and pedagogical applications, the corpus can support studies of musical similarity and influence through the structural patterns shared among artists, contributing to the mapping of stylistic relationships across the MPB tradition. Such representations may also form the basis for recommendation systems that rely on musical content rather than audio-based features alone. Finally, the dataset opens broader cultural and sociomusicological investigations: the study of rhythmic and melodic diversity within MPB in relation to other global repertoires, and analyses of how stylistic features relate to historical periods and cultural movements. 

Together, these possibilities position the MPB Corpus and the proposed analytical models as versatile resources for a wide range of inquiries across musicology, MIR, pedagogy, and cultural analysis.

\section{Corpus development and future expansion}
\label{sec:corpus-development}
The present dataset represents the culmination of approximately five years of work, from the initial conception of the project to the completion of its first public release, an effort extending far beyond the annotation of the musical pieces themselves. A substantial part of the project was devoted to developing, refining, and validating the analytical methodology, an iterative process involving multiple cycles of hypothesis formulation, implementation, testing, and revision. Further effort went into the construction and analysis of the Tom Jobim corpus, which served as the basis for the analytical models described in \cite{almada-harmonia-jobim} and \cite{almada-melodia-jobim}, and into the preparation and publication of these reference volumes.

Publishing both the analytical methodology and the present dataset represents an important milestone for the project. By documenting the analytical procedures in detail in English and making the resulting corpus publicly available, the project establishes a stable methodological framework that can be more easily disseminated, reproduced, and adopted by other researchers. Most procedures have now been fully formalized, although work is still in progress on a rigorous formalization of melodic segmentation, currently the least standardized stage of the analytical pipeline. Completing this step should further improve reproducibility and facilitate the training of new collaborators.

The long-term objective of the project is to expand the corpus to approximately fifty composers representing a broader range of Brazilian popular music, and this expansion is already underway. Now that the analytical workflow has been consolidated and documented, additional researchers have been trained to apply the proposed methodology, substantially increasing the project's annotation capacity. The effort required to extend the corpus is therefore no longer expected to scale linearly with the time invested in the first release, much of which was dedicated to establishing the analytical framework itself.

One of the main remaining challenges concerns composers for whom reliable published scores are unavailable. In such cases, the analytical workflow must be preceded by manual transcription of the musical pieces directly from audio recordings, a time-consuming process that remains difficult to automate at the level of accuracy the project requires. Nevertheless, the larger annotation team and the consolidated analytical methodology provide favorable conditions for the continued growth of the corpus.

In parallel with the inclusion of additional composers, work is nearing completion on a control corpus comprising representative works from \textit{choro}, \textit{samba}, and jazz, which will provide an important reference for future comparative studies involving Brazilian popular music and related musical traditions.

Taken together, these developments are expected to substantially accelerate future releases of the corpus while maintaining the consistency and quality standards here established. Achieving the long-term goal of approximately fifty annotated composers will provide a unique resource for large-scale quantitative investigations of Brazilian popular music, enabling systematic studies of stylistic similarity, historical evolution, genre interaction, and computational modeling that are currently impractical due to the lack of a sufficiently broad and consistently annotated corpus.

\section{Conclusion}
\label{sec:conclusion}
In this work, we present the MPB Corpus, the most comprehensive dataset dedicated to computational musicology of Brazilian music. The corpus is based on the analysis of melodic contour, melodic rhythm, harmony, and the relationship between melody and harmony for 500 songs, 50 for each of ten prominent figures within the context of MPB: Tom Jobim, Caetano Veloso, Edu Lobo, Chico Buarque, Milton Nascimento, Ivan Lins, Gilberto Gil, Djavan, Rita Lee, and João Bosco. Given the scope of the work, specific analytical models had to be developed and are also presented here -- the Genera of Chord Types and the Melodic Filtering Model -- together with specific metrics and graphical visualizations for the musical parameters under study. We also provided a brief exploratory analysis of the collected data, complemented by a permutation-based statistical test, illustrating the potential of the dataset to formulate and systematically answer musicological questions about MPB, and outlined potential applications alongside the construction effort behind the corpus, ongoing methodological refinements, and plans for its long-term expansion. The following paragraphs outline some of the most promising lines of future work already under development.

MPB is strongly influenced by genres such as \textit{samba}, \textit{choro}, and jazz. To support comparative analyses, a complementary control corpus of 150 songs (50 from each genre) is being finalized, providing reference points for clustering and other quantitative studies of stylistic influence. In parallel, the main dataset will expand beyond the ten composers included in this first phase, toward the long-term goal of approximately fifty. For some artists of interest (e.g., Cartola, Dori Caymmi, Joyce), reliable scores are limited, requiring manual transcription and resulting in smaller samples of around 20-30 pieces; preliminary bootstrap analyses suggest that such sample sizes remain sufficient to capture each composer’s core stylistic variability. Finally, given the large number of collected musical attributes, interactive dashboards are currently being developed to facilitate exploratory data analysis, making the dataset accessible not only to MIR researchers but also to a broader audience with general musical interests.

Clustering analysis offers a systematic way to identify similarities and contrasts among composers, either by grouping them on bar-chart distributions of selected musical parameters or by treating each piece as a ``bag-of-words''~consisting of its r-words, c-words, note-functions, and functional categories (or any combination of these aspects), allowing the use of text-clustering methods such as Latent Dirichlet Allocation. Although this approach disregards syntactic structure, preliminary results suggest that it is still effective in revealing stylistic affinities and influences; future work will extend the framework to incorporate sequential relationships. Beyond analytical applications, the dataset also serves as a generative resource for composition, enabling the replication of binary relationships between functional categories or the creation of new material via Markov chains, providing composers with stylistically informed musical material while preserving space for individual creativity; dedicated software to support these applications is currently under development.

By providing openly accessible annotations, documentation, and code, the project contributes to a more diverse and reusable music-data ecosystem. The corpus also illustrates how expert symbolic annotations derived from copyright-protected materials can be shared responsibly while making their provenance, scope, and limitations explicit. Its versioned release supports long-term reuse, citation, and future extension by the MIR community.

We hope that the MPB Corpus will have a positive impact on the musicological community, particularly by shedding light on the rich and multifaceted tradition that is Brazilian popular music, which nonetheless remains largely unexplored in digital musicology. 

%%%%%%%%%%%%%%%%%%%%%%%%%%%%%%%%%%%%%%%%%%%%%%%%%%%%%%%%%%%%%%%%%%%%%%%%%%%%%%%%
% Please do not touch.
% Print Endnotes
\IfFileExists{\jobname.ent}{
   \theendnotes
}{
   %no endnotes
}
%%%%%%%%%%%%%%%%%%%%%%%%%%%%%%%%%%%%%%%%%%%%%%%%%%%%%%%%%%%%%%%%%%%%%%%%%%%%%%%%

\section*{Acknowledgements}
We would like to thank Luiz W. P. Biscainho for proofreading a preliminary version of this paper; Zico B. de F. Costa for assistance with the code for metadata retrieval; the reviewers and editors for their valuable and constructive feedback, which contributed to improving this work; and the ISMIR Society for the APC waiver.

\section*{Competing Interests}
The authors have no competing interests to declare.

\section*{Data Accessibility}
To foster reproducibility and further research, we provide access to the MPB Corpus and related resources.
\begin{itemize}
    \item Dataset: The MPB Corpus is available under a CC-BY-4.0 license on Zenodo: \url{https://zenodo.org/records/22737013}.

    \item Code and supplementary material: Metadata, supplementary material, documentation, and accompanying code for reproducibility are available on GitHub: \url{https://github.com/ProjetoMPB/mpb-corpus}.
\end{itemize}

\section*{Authors' contributions}
CLA conceptualized the research, developed the annotation methodologies, performed the data annotation, reviewed the dataset, and contributed to the writing (draft and final versions). HTC contributed to the conceptualization, developed code for statistical analysis, conducted statistical analysis, visualization, reviewed the dataset, and led the writing (draft and final). FDM contributed to the software, data curation, dataset review, and writing (draft).

\section*{Generative AI Use Statement}
We used generative AI tools to assist with text review, repository and data organization, and, in some instances, code generation. All AI-generated outputs were carefully reviewed by the authors and verified for correctness and appropriateness before inclusion in the final manuscript, repository, or codebase. The authors take full responsibility for the content of this work.

%%%%%%%%%%%%%%%%%%%%%%%%%%%%%%%%%%%%%%%%%%%%%%%%%%%%%%%%%%%%%%%%%%%%%%%%%%%%%%%%
% Bibliography
%%%%%%%%%%%%%%%%%%%%%%%%%%%%%%%%%%%%%%%%%%%%%%%%%%%%%%%%%%%%%%%%%%%%%%%%%%%%%%%%

% For bibtex users:
\bibliography{refs}

\end{document}